\documentclass[lineno]{jfm}
\usepackage{ulem}
\usepackage{newtxtext}
\usepackage{newtxmath}
\usepackage{natbib}
\usepackage{tikz}
\usepackage{todonotes}
\usepackage{amsmath}
\usepackage{hyperref}
\hypersetup{
    colorlinks = true,
    urlcolor   = blue,
    citecolor  = black,
}

\newcommand{\RomanNumeralCaps}[1]
\linenumbers
\usepackage{gensymb}
\usetikzlibrary{patterns.meta}
\usepackage{color}
\usepackage{subcaption}
\usepackage{graphicx}
\definecolor{green}{RGB}{34,150,34}
\definecolor{maroon}{rgb}{0.76, 0.13, 0.28}

\title{Stability analysis of time-periodic shear flow generated by an oscillating density interface}
\shorttitle{Stability of time-periodic shear flow generated by an oscillating density interface}
 \author{Lima Biswas\aff{1}
  \and Anirban Guha\aff{2}
  \corresp{\email{anirbanguha.ubc@gmail.com}}}
 \affiliation{\aff{1} Department of Mathematics, Gandhi Institute of Technology and Management (GITAM), Hyderabad 502329, India.
 \aff{2} School of Science and Engineering, University of Dundee, Dundee DD1 4HN, UK.}
\begin{document}
\maketitle

\begin{abstract}
We consider the conceptual two-layered oscillating tank of \cite{inoue2009efficiency}, which mimics the time-periodic parallel shear flow generated by low-frequency (e.g. semi-diurnal tides) and small-angle oscillations of the density interface. Such self-induced shear of an oscillating pycnocline may provide an alternate pathway to pycnocline turbulence and diapycnal mixing in addition to the turbulence and mixing driven by wind-induced shear of the surface mixed layer. We theoretically investigate shear instabilities arising in the inviscid two-layered oscillating tank configuration and show that the equation governing the evolution of linear perturbations on the density interface is a Schr\"{o}dinger-type ordinary differential equation with a periodic potential. The necessary and sufficient stability condition is {governed by a nondimensional parameter $\beta$ resembling the inverse Richardson number}; for two layers of equal thickness, instability arises when $\beta\!>\!1/4$. When this condition is satisfied, the flow is initially stable but finally tunnels into the unstable region after reaching the time marking the turning point. Once unstable, perturbations grow exponentially and reveal characteristics of Kelvin-Helmholtz (KH) instability.  The Modified Airy Function method, which is an improved variant of the  Wentzel–Kramers–Brillouin (WKB) theory, is implemented to obtain a uniformly valid, composite approximate solution to the interface evolution.  Next, we analyse the fully nonlinear stages of interface evolution by modifying the circulation evolution equation in the standard vortex blob method, which reveals that the interface rolls up into KH billows.  Finally, we undertake real case studies of Lake Geneva and Chesapeake Bay to provide a physical perspective.

\end{abstract}


\section{Introduction}

Turbulent mixing in the stratified ocean interior, originating from internal gravity wave breaking \citep{mackinnon2017climate}, plays a central role in determining the global distributions of carbon, heat, nutrients, plankton, and other salient tracers \citep{wunsch2004vertical}. Kelvin-Helmholtz (KH) instability is arguably one of the key mechanisms generating such small-scale turbulence and diapycnal mixing \citep{dauxois2021confronting,caulfield2021layering}. Wind-induced shear in the surface mixed layer can render the internal (or interfacial) gravity waves at the pycnocline unstable, causing the interface to roll up into billow-like structures -- the hallmark of  Kelvin-Helmholtz (KH) instability \citep{smyth2012ocean}. This paper digresses from the typical studies on KH instability where the background shear flow is assumed to be \textit{steady}, for example, see \cite{caulfield2000anatomy,mashayek2012zoo,rahmani2014effect,mashayek2021goldilocks}.
%, which is often a sensible approximation for the surface mixed layer.
Instead, our study is motivated by settings where the background shear has an \textit{explicit time dependence}, periodic to be specific.  During the summer seasons, shelf seas \citep{mihanovic2009diurnal,guihou2018kilometric}, lakes \citep{lemmin2005internal} and estuaries \citep{sanford1990} often display a  two-layered stratification; moreover, the pycnocline can undergo slow basin-scale (low-mode) oscillations. Such `see-saw' oscillations of the pycnocline can generate a time-periodic buoyancy-driven shear flow, which induces mixing, as revealed in field observations    \citep{sanford1990}.
%While pycnocline oscillations are known to drive mixing \citep{sanford1990}, the fundamental mechanisms at play are little understood.
Within the shelf sea seasonal pycnocline, fine-scale shear
and stratification observations suggest a state of marginal
stability \citep{mackinnon2005spring} such that the addition of even a
small amount of shear due to pycnocline oscillations could
be sufficient to trigger the cascade of energy from the low
mode oscillations, such as inertial oscillations and internal
tides, into pycnocline turbulence, and thence to enhanced
levels of mixing. The aim of this paper is to understand this alternate pathway to instability arising from an oscillating pycnocline. 

 By titling a tank consisting of a two-layered fluid at a constant angle,  \cite{thorpe1969experiments} generated a constantly accelerating buoyancy-driven shear flow, and supported the experimental observations on the ensuing KH instabilities using theoretical arguments. Thorpe's tilting tank setup has paved the pathway for many controlled experimental and numerical studies on stratified shear instabilities induced by sloping density interface, and remains an area of active research \citep{atoufi2023stratified,zhu2024long}. More complex density interface profiles in controlled two-layered setups have also been studied in recent years. One such example is the generation of internal solitary waves at the density interface, whose subsequent breaking leads to KH instabilities \citep{carr2017characteristics}. 
 Thorpe's setup was numerically extended to a two-layered oscillating tank by \cite{inoue2009efficiency}, and the same setup has been recently studied by \cite{lewin2022stratified}. The two-layered oscillating tank configuration provides a simplified analogue of an oscillating pycnocline; while the analogy with standing-wave oscillations (e.g. seiches) is obvious, a superficial connection can be made with pycnocline oscillations induced by travelling, long interfacial waves, which induce a near-parallel, oscillating shear flow.  Indeed, the Direct Numerical Simulation (DNS) studies of \cite{inoue2009efficiency} and \cite{lewin2022stratified} have clearly revealed how a slowly oscillating density interface generates time-periodic shear flow that renders the interface unstable, and leads to the formation of KH billows, which subsequently drive turbulence and mixing. Their studies may provide an explanation of the appearance of a long train of KH billows superposed on internal gravity waves of lower frequency (corresponding to semi-diurnal tides) at about $560$ m depth over the Great Meteor Seamount \citep{van2010deep}. Unlike the DNS studies of \cite{inoue2009efficiency} and \cite{lewin2022stratified}, which have focused on the `late-time dynamics' like turbulence and mixing,  the onset of instability is the cornerstone of our analysis (although we also analyse the initial nonlinear stages of billow formation until saturation). As highlighted in \cite{lewin2022stratified}, the time of saturation of the KH billow relative to the time period of the background shear plays a central role in shaping the fully nonlinear stages of turbulent breakdown of the billow and therefore crucially affects the energetics and mixing. Since the standard linear stability analysis of stratified shear flows \citep{drazin2004} is based on steady background conditions, it is unlikely to capture some of the key properties of instabilities arising in background flows with periodic time dependence. This necessitates a linear stability analysis of non-autonomous systems. 
 %The numerical two-layered oscillating tank setup (see figure \ref{fig:schematic}), first introduced by \cite{inoue2009efficiency} and recently studied by \cite{lewin2022stratified}, provides a simplified analogue of an oscillating pycnocline, and has successfully captured the formation of KH billows.  This provides us with the justification to further investigate the hydrodynamic instabilities arising in the above-mentioned conceptual setup.  However, unlike the Direct Numerical Simulation (DNS) studies of \cite{inoue2009efficiency} and \cite{lewin2022stratified}, which focus on the highly nonlinear stages of KH-induced turbulence and diapycnal mixing, 
  Another key feature of this problem is the large separation of time scales -- high-frequency interfacial waves (which are rendered unstable) being forced by low-frequency density interface oscillations. The numerical simulations of \cite{inoue2009efficiency} and \cite{lewin2022stratified} do not indicate any finely tuned frequency ratios for which the instability occurs. This implies that parametric instability, for example, the problem studied in \cite{kelly1965stability}, is unlikely to be the driving mechanism.  Along similar lines, the Generalised stability theory for non-autonomous systems \citep{farrell1996generalized}
might not be best suited for the purpose of this study. This necessitates a bespoke analysis starting from the fundamental principles.

The paper is organised as follows. \S \ref{sec:gov-eq} discusses the governing equations for the background flow and adds linear perturbations to the system to derive the equation governing the interfacial dynamics. The condition for instability is also obtained.  \S \ref{sec:linstab} deals with various kinds of linear stability analyses, e.g.\,Floquet theory, normal mode theory (based on steady-state analysis) and Modified Airy Function (MAF) technique, and provides comparisons with numerical solutions. Building on the current vortex methods technique, \S \ref{sec:numer} incorporates unsteady effects and captures the nonlinear evolution of the interface. Physical interpretation of the system is provided in  \S \ref{bsec:case_study} by undertaking two case studies -- Lake Geneva and Chesapeake Bay. Finally, the paper is summarised and concluded in \S \ref{sec:disc_conc}.
\section{Governing equations}
\label{sec:gov-eq}

\begin{figure}
    \centering
    \includegraphics[width=0.8\linewidth]{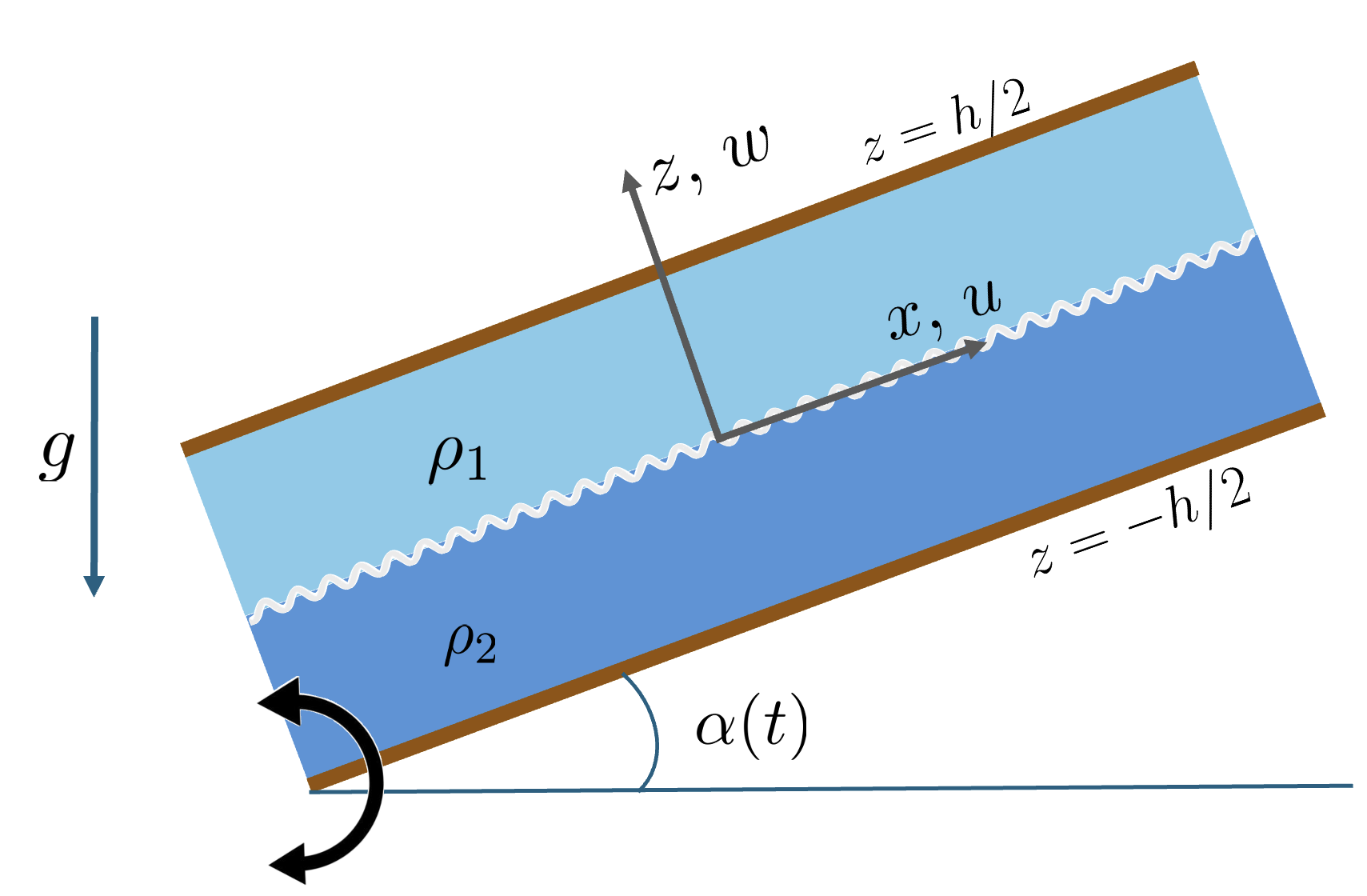}
    \caption{Schematic diagram of a two-layered oscillating tank, which can render short interfacial gravity waves unstable.}
    \label{fig:schematic}
\end{figure}
% \begin{figure}\label{fig:schematic}
%     \centering
% \begin{tikzpicture}
%     % Draw the tilted tube
%     \draw[->, thick] (-5,-1) -- (3,0.6)node[anchor=north] {$u, x$};
%     \draw[thick] (-5,-2.5) -- (3,-0.9) node[anchor=north] {$z=-h/2$};
%      \draw[thick] (-5,0.5) -- (2.5,2)node[anchor=north] {$z=h/2$};
    
%     % Draw the boundaries of the separated fluids
%      \draw[thick] (-3,-2.1) -- (2,-2.1);

%     % Draw z-axis and labels

%           \draw[->, thick] (-2.71,-2.042) -- (-3.4,1.4)node[anchor=west] {$w, z$};

% \draw[<->,line width=0.5mm] (-3.3,-0.94) arc[start angle=-130, end angle=165, radius=0.5];

%     % Draw g vector
%     \draw[->, thick] (-6,-0.5) -- (-6,-1.5) node[anchor=west] {$g$};
    
%     % Draw density labels
%     \node at (-4,-0.5) {$\rho_1$};
%     \node at (-4,-2) {$\rho_2$};

%     % Draw angle alpha
%     \draw[<->] (0,-2.1) arc[start angle=0, end angle=32, radius=1] node[midway, anchor=west] {$\alpha(t)$};

%         \foreach \x in {-5, -4.8, ..., 2.5} {
%         \draw (\x,0.2*\x+1.5) -- (\x-0.1, 0.2*\x+1.6);
%     }

%     \foreach \x in {-5, -4.8, ...,3} {
%         \draw (\x,0.2*\x-1.5) -- (\x-0.1, 0.2*\x-1.6);
%     }

% \end{tikzpicture}
% \caption{Schematic diagram of the two-layered oscillating channel flow.}
% \label{fig:schematic}
% \end{figure}

\subsection{Background flow} 
\label{subsec:bg_flow}
 We consider the two-dimensional (2D) flow produced by oscillating a horizontal tank having a rectangular cross-section of depth $h$, while the along-channel ($x$) and cross-channel ($y$) dimensions are infinite. The $z$ direction is positive upward; see figure \ref{fig:schematic} for the schematic of the setup. 
 The tank consists of a Boussinesq, two-layered, inviscid and immiscible fluid with the following density distribution:
\begin{align}\label{eqn:rho}
    \rho(z) =
    \left\{
    \begin{aligned}
    & \rho_1,~~~ 0<z<h/2 \\
    & \rho_2,~~~ -h/2<z<0.
    \end{aligned}
    \right.
\end{align}
The fluid is stably stratified, i.e. $\rho_1\!<\!\rho_2$. A more generic setting in which the two layers have unequal depths is discussed in Appendix \ref{appA}.
Throughout the paper,  subscripts $1$ and $2$ will respectively denote quantities associated with the upper and lower ﬂuid layers.
%The angle of oscillation $\alpha(t)$ is kept small
%obeying \,$\max |\alpha(t)| \lesssim 5^\circ$.
The system is restricted to a small angle of oscillation  $\alpha(t)$, typically $\max |\alpha(t)|\!\lesssim\!7^\circ$. The small tilt of the density interface functions as a geophysically plausible mechanism for generating shear; the vertical motions induced by the sloping interface have a negligible effect on the fluid dynamics 
\citep{lewin2022stratified}.
%We consider the two-dimensional rectangular co-ordinate system with the $x$-axis  aligned along the plane of the highest slope $\alpha(t)$ and $z$-axis  perpendicular to it, as illustrated in the schematic diagram~\ref{fig:schematic}. For simplicity, we assume the fluids in both the layers to be homogeneous,   inviscid and incompressible. Further, the eﬀect of surface tension is neglected.
For the setting in figure \ref{fig:schematic}, the governing Navier-Stokes  equations  under the Boussinesq approximation read \citep{inoue2009efficiency,lewin2022stratified}
\begin{subequations} \label{eqn:main_equation}
 \begin{align}
     % \begin{aligned}
    \frac{\mathrm{D}u}{\mathrm{D}t} = -\frac{1}{\rho_0}\frac{\partial p}{\partial x} - g\frac{\rho}{\rho_0}\sin \alpha(t) + 2 \alpha_{,t} w, \label{eqn:main_equation_u} \\
     \frac{\mathrm{D}w}{\mathrm{D}t} = -\frac{1}{\rho_0}\frac{\partial p}{\partial z} - g\frac{\rho}{\rho_0}\cos \alpha(t) - 2 \alpha_{,t} u \label{eqn:main_equation_w}.
% \end{aligned}
 \end{align}
 \end{subequations}
 % \todo{g is a scalar, so don't make it bold}
Here $\mathrm{D}/\mathrm{D}t
= \partial/\partial t+\mathbf{u}\cdot\nabla$, where $\nabla=(\hat{\mathbf{x}}\,\partial/\partial x,\,\hat{\mathbf{z}}\,\partial/\partial z)$, $\mathbf{u} = \left(u,\, w\right)$ is the velocity vector, $\rho_0$ is the constant reference density, $g$ is the acceleration due to gravity, and the suffix `$,t$' denotes ordinary derivative $d/dt$. The last terms in \eqref{eqn:main_equation_u}-\eqref{eqn:main_equation_w} represent  Coriolis accelerations arising from the time-dependent rotation. The incompressibility condition is satisfied by the continuity equation
\begin{equation}
    \frac{\partial u}{\partial x}+\frac{\partial w}{\partial z}=0.
    \label{eq:cont}
\end{equation}

%\subsection{Background flow}
The background state velocity is assumed to be parallel to the $x$-axis, i.e. $\mathbf{U}=(U,0)$. Hence the continuity equation \eqref{eq:cont} implies that $U$ is only a function of $z$ and $t$. Therefore, the background state equations yield 
%The time-dependent background state velocity $(U(z,t),~0)$ and the pressure $P(x,z,t)$ satisfy 
\begin{subequations}
    \begin{align}
    & \frac{\partial U}{\partial t} = -\frac{1}{\rho_0}\frac{\partial P}{\partial x} - g\frac{\rho}{\rho_0}\sin \alpha (t), \label{eqn:U equation}\\
     & P = - \int \left[g \rho(z) \cos \alpha (t) + 2 \rho_0 \alpha_{,t} U(z, t)\right] \mathrm{d}z + \mathcal{G}(x,t),\label{eqn:P equation}
\end{align}
\end{subequations}
where $P(x,z,t)$ denotes the background state pressure.
%We note in passing that the pressure gradient term in \eqref{eqn:U equation} is omitted  in 
%\citet[eq.\,(7)]{inoue2009efficiency} and \citet[eq.\,(2.3)]{lewin2022stratified}.

Equation~\eqref{eqn:rho} 
% \todo{simply write (2.3), and not equation (2.3). The only place where you should write the word ``equation'' before equation number is if you begin a sentence with an equation}
indicates that $\rho$ is a function of $z$ only. Consequently from \eqref{eqn:P equation}, $P(x,z,t)$ can be expressed as the sum of a function that is dependent on $z$  and $t$, and another function $\mathcal{G}$ dependent on $x$ and $t$. This implies $\partial P/\partial x$ is a function of $x$ and $t$. However, as per \eqref{eqn:U equation}, both
 $\partial U/\partial t$ and $ g({\rho}/{\rho_0})\sin \alpha(t)$ are functions of $z$ and $t$ only. Therefore, $\partial P/\partial x$ must be a function of $t$ alone. Since the tank is enclosed and there is no net flux across the walls, we must have \citep{thorpe1969experiments}
\begin{align}\label{eqn:zero_net_flux}
     \int_{-h/2}^{h/2} U(z, t)\, \mathrm{d}z =0.
\end{align}
The above condition can be used to obtain the background state velocity profile. In this regard, we integrate
 \eqref{eqn:U equation} in both $z$ and $t$ and substitute  \eqref{eqn:zero_net_flux}:
%Combining \eqref{eqn:zero_net_flux} and \eqref{eqn:U equation} we obtain
\begin{equation}
    \int_0^t \frac{\partial P}{\partial x} \mathrm{d}t = -\frac{g}{h} \int_{-h/2}^{h/2} \rho \mathrm{d}z \int_0^t\sin\alpha(\tilde{t})\, \mathrm{d}\tilde{t}.
    \label{eq:press}
\end{equation}
Finally, we integrate \eqref{eqn:U equation} in $t$  and combine with equations  \eqref{eq:press} and \eqref{eqn:rho}, which  yields the background state velocity profile: 
%The above equation is substituted back in \eqref{eqn:U equation}, integrated in time, and combined with  \eqref{eqn:rho} to yield the background state velocity profile:
%Substituting $U(z,t)$ from~\eqref{eqn:U equation} into \eqref{eqn:zero_net_flux} and using~\eqref{eqn:rho} we have
\begin{align} 
    U(z,t) =\left\{
    \begin{aligned}
       &  U_1(t)=\frac{g^\prime}{2}
     \int_0^t \sin \alpha(\tilde{t}) \, \mathrm{d}\tilde{t}, \,\,\,\,\,\,\,\,\, 0<z<h/2 \\
    &  U_2( t)=- \frac{g^\prime}{2}  \int_0^t \sin \alpha(\tilde{t}) \, \mathrm{d}\tilde{t}, \,\,\,\,\,\,-h/2<z<0, 
    \end{aligned}
\right.
\label{eq:base_vel}
    \end{align}
where
% \begin{equation} \label{eqn:U_1 and U_2}
% \begin{aligned}
%     U_1(t) = \frac{g^\prime}{2}
%      \int_0^t \sin \alpha(t) \, \mathrm{d}t, \quad
%     U_2(t) = - \frac{g^\prime}{2}  \int_0^t \sin \alpha(t) \, \mathrm{d}t,  \quad \text{and} \quad  g^\prime=\frac{g}{\rho_0}(\rho_2-\rho_1)
% \end{aligned}
% \end{equation}
 $g^\prime\!=\!{g}(\rho_2-\rho_1)/{\rho_0}$ is the reduced gravity. We  assume that the tilt angle varies sinusoidally with time:  
\begin{align}\label{eqn:alpha}
    \alpha(t)=\alpha_f\sin(\omega_f t),
\end{align}
where $\alpha_f$ and $\omega_f$  respectively denote the maximum angular amplitude and frequency of oscillation of the tilted tank. Since we have restricted our study to small oscillations ($\alpha_f\!\ll\!1$), this implies $\sin \alpha \approx \alpha$ and $\cos \alpha \approx 1-\alpha^2/2$.
%Further, we assume that the angle of oscillation $\alpha(t)$ is small, so that $\sin \alpha \approx \alpha$ and $\cos \alpha \approx 1-\alpha^2/2$.
Under this assumption, the background state velocity in \eqref{eq:base_vel} simplifies to

\begin{subequations}
\begin{align}
        U_1(t)&= 
 %  \frac{g^\prime}{2} \int_0^t a\sin(\omega_f t) \mathrm{d}t = 
    \frac{g^\prime}{2} \frac{\alpha_f}{\omega_f} \bigg[1-\cos\left(\omega_f t\right)\bigg],\,\,\,\,\,\,\,\,\, 0<z<h/2  \label{eq:base_vel_simp_a} \\ \quad  U_2(t)& = - \frac{g^\prime}{2} \frac{\alpha_f}{\omega_f} \bigg[1-\cos\left(\omega_f t\right)\bigg], \,\,\,\,\, -h/2<z<0.\label{eq:base_vel_simp_b}
    \end{align}
\end{subequations}
The fact that the signs in \eqref{eq:base_vel_simp_a}--\eqref{eq:base_vel_simp_b} agree with our intuition can be verified by simply inspecting figure \ref{fig:schematic}, which can be regarded as the maximum angle in the counter-clockwise direction attained by the tank,  equaling $\alpha_f$. As per \eqref{eqn:alpha} this implies $\omega_f t\!=\!\pi/2$, which, when substituted into \eqref{eq:base_vel_simp_a}--\eqref{eq:base_vel_simp_b}, yields $U_1\!=\!-U_2\!=\!g'\alpha_f/(2\omega_f)$. 
The signs make sense because the heavier, lower layer must have a negative velocity since it would flow downwards due to gravity, and the upper layer should compensate for this by flowing upwards, yielding a positive velocity.
%The discrepancy between the signs in \eqref{eq:base_vel_simp_a}--\eqref{eq:base_vel_simp_b} and that in \citet[eq.\,(11)]{inoue2009efficiency} arises because the authors have neglected the pressure gradient term  in \eqref{eqn:U equation}.
%Note that the above background flow profile and that of \citet[eq.\,(11)]{inoue2009efficiency} have opposite signs. 

\subsection{Linear perturbations}
\label{subsec:lin_per}
We now consider the effect of infinitesimal perturbations added to the background state, and follow an approach similar to \cite{thorpe1969experiments}.
The displacement of the two-fluid interface, given by $z\!=\!\xi(x,\,t)$, sets up an irrotational perturbation velocity field in the two layers. The total velocity reads 
\begin{align}\label{eqn:perturbed_velocity}
\mathbf{u}= \left\{ 
    \begin{aligned}
           & U_1(t) \hat{\mathbf{x}} +\nabla \phi_1(x,z,t), \qquad 0 <z<h/2 \\
      &  U_2(t) \hat{\mathbf{x}} +\nabla \phi_2(x,z,t), \qquad -h/2<z< 0,
    \end{aligned}
    \right.
\end{align}
where $\phi_{1}$ and $\phi_2$ are respectively the perturbation velocity potentials in the upper and lower fluid layers, which satisfy Laplace's equation in their respective layers:
\begin{subequations}
    \begin{align}
    \nabla^2 \phi_1&=0, \qquad 0 <z<h/2 \label{eq:Lap_upper} \\
    \nabla^2 \phi_2&=0, \qquad  -h/2<z< 0. \label{eq:Lap_lower}
\end{align}
\end{subequations}
% Likewise, the total pressure can be expressed as
% \begin{align}\label{eqn:perturbed_pressure}
% p= \left\{ 
%     \begin{aligned}
%            & P(t)  +p_1(x, z,\,t), \qquad \xi(x,\,t) <z<h/2 \\
%       &  P(t)  +p_2(x, z,\,t), \qquad -h/2<z< \xi(x,\,t) 
%     \end{aligned}
%     \right. 
% \end{align}
% where  $p_{1}$ and $p_2$ are respectively the perturbation pressure in the upper and lower layers.

The linearised dynamical boundary conditions at $z=0$ for the two layers are as follows: 
\begin{subequations}
    \begin{align}
    \frac{P_1}{\rho_1}+\frac{\partial \phi_1}{\partial t} + \frac{\partial }{\partial t}\left(U_1 x\right) + U_1 \frac{\partial \phi_1}{\partial x} + gx \sin\alpha(t) + gz\cos \alpha(t)=A_1(t),\label{eqn:dynamic_bc_1}\\
     \frac{P_2}{\rho_2}+\frac{\partial \phi_2}{\partial t} + \frac{\partial }{\partial t}\left(U_2 x\right) + U_2 \frac{\partial \phi_2}{\partial x} + gx \sin\alpha(t)+ gz\cos \alpha(t) =A_2(t),\label{eqn:dynamic_bc_2}
\end{align}
\end{subequations}
where $P_1$ and $P_2$ respectively denote pressure in the upper and lower fluid layers, and $A_1$ and $A_2$ are arbitrary functions of time.
% Continuity of pressure across the density interface allows equating the linearised dynamic boundary condition above and below the interface, yielding
% \begin{equation}\label{eqn:lin_bernoulli}
%     \begin{aligned}
%    \frac{\partial \phi_1}{\partial t}-\frac{\partial \phi_2}{\partial t} +  U_1 \frac{\partial \phi_1}{\partial x} -  U_2 \frac{\partial \phi_2}{\partial x}=g'\xi \cos \alpha(t) \,\,\, \quad \text{at} \quad z=0.
% \end{aligned}
% \end{equation}
Likewise, the linearised kinematic boundary conditions at $z=0$ for the two layers read
\begin{subequations}\label{eqn:kinematic_bc}
    \begin{align} 
    \frac{\partial \xi}{\partial t} + U_1(t)
      \frac{\partial \xi}{\partial x} 
    = \frac{\partial \phi_1}{\partial z}, \label{eqn:kinematic_bc_1} \\
       \frac{\partial \xi}{\partial t} + U_2(t)  \frac{\partial \xi}{\partial x}       
       = \frac{\partial \phi_2}{\partial z}. \label{eqn:kinematic_bc_2}
\end{align}
\end{subequations}
 The upper and lower boundaries satisfy the impermeability condition: 
 \begin{subequations}
     \begin{align}
   \frac{\partial \phi_1}{\partial z} &= 0 \quad \text{at} \quad z=h/2, \label{eqn:imper_bc_1}  \\
     \frac{\partial \phi_2}{\partial z}& = 0 \quad \text{at} \quad z=-h/2.\label{eqn:imper_bc_2}
\end{align}
 \end{subequations}

Next, we assume the perturbations in the form of Fourier modes: 
\begin{subequations}\label{eqn:fourier_mode}
    \begin{align}
    &\xi(x,t)=  \eta(t) e^{\mathrm{i}kx}, \label{eqn:fourier_mode_xi}\\
   & \phi_1(x,z,t) =\psi_1(t)\cosh k\left(z-\frac{h}{2}\right)e^{\mathrm{i}kx}, \label{eqn:fourier_mode_phi1}\\
   & \phi_2(x,z,t) =\psi_2(t)\cosh k\left(z+\frac{h}{2}\right)e^{\mathrm{i}kx}, \label{eqn:fourier_mode_phi2}
\end{align}
\end{subequations}
where $k$ denotes the wavenumber and it is understood that the real parts of terms appearing on the right-hand sides are to be taken. It is straightforward to verify that the perturbation velocity potentials satisfy Laplace's equation \eqref{eq:Lap_upper}--\eqref{eq:Lap_lower} and the impermeability condition \eqref{eqn:imper_bc_1}--\eqref{eqn:imper_bc_2}. The continuity of pressure across the interface  ($P_1\!=\!P_2$) is applied in \eqref{eqn:dynamic_bc_1}--\eqref{eqn:dynamic_bc_2}, thereby combining these two equations into one:
\begin{align}
   & \rho_1 A_1 -\rho_1 \psi_{1,t} \cosh \left(\frac{kh}{2}\right) e^{\mathrm{i}kx} - \rho_1 x U_{1,t} -\mathrm{i}k \rho_1 U_1 \psi_1 \cosh \left(\frac{kh}{2}\right) e^{\mathrm{i}kx} - \rho_1 gx\sin \alpha(t) \nonumber\\ &    -  \rho_1 g \eta \cos \alpha(t)  e^{\mathrm{i}kx} 
     = \rho_2 A_2 -\rho_2 \psi_{2,t} \cosh \left(\frac{kh}{2}\right) e^{\mathrm{i}kx} - \rho_2 x U_{2,t} -\mathrm{i}k \rho_2 U_2 \psi_2 \cosh \left(\frac{kh}{2}\right) e^{\mathrm{i}kx}  \nonumber \\ &  - \rho_2 gx\sin \alpha(t) -  \rho_2 g \eta \cos \alpha(t)  e^{\mathrm{i}kx}. 
\end{align}
Note that in the above equation, we have substituted  \eqref{eqn:fourier_mode_xi}--\eqref{eqn:fourier_mode_phi2}. Next we collect the coefficients of $e^{\mathrm{i}kx}$
\cite[Appendix A]{thorpe1969experiments}, and invoke the Boussinesq approximation:
%we substitute \eqref{eqn:fourier_mode_xi}--\eqref{eqn:fourier_mode_phi2}, and collect the coefficient of $e^{\mathrm{i}kx}$. 
 %This yields      We follow the procedure detailed in \citet[Appendix A]{thorpe1969experiments} and substitute the expressions  \eqref{eqn:fourier_mode_xi}--\eqref{eqn:fourier_mode_phi2} in \eqref{eqn:dynamic_bc_1}. Collecting the coefficients of the $e^{\mathrm{i}kx}$ we obtain 
% . The expressions in \eqref{eqn:fourier_mode_xi}--\eqref{eqn:fourier_mode_phi2} are substituted in \eqref{eqn:lin_bernoulli} and \eqref{eqn:kinematic_bc_1}--\eqref{eqn:kinematic_bc_2}, and these equations are then combined to yield
\begin{align}
     \left(\psi_{1,t}-\psi_{2,t}\right)\cosh\left(\frac{kh}{2}\right) + \mathrm{i}k \bigg(  U_1  \psi_1 -  U_2  \psi_2 \bigg)\cosh \left(\frac{kh}{2}\right) 
   =     g' \eta \cos \alpha(t). 
\end{align}
Finally, we substitute the linearised kinematic boundary conditions \eqref{eqn:kinematic_bc_1}--\eqref{eqn:kinematic_bc_2} in the above equation, yielding a second-order ordinary differential equation (ODE) only in terms of $\eta$:
\begin{align}\label{eqn:eta_gen}
   \eta_{,tt} +\bigg[ \underbrace{\frac{kg^\prime}{2} \tanh\bigg(\frac{kh}{2}\bigg)}_{\text{$\omega^2$}} \cos  \alpha(t) -\frac{k^2}{2}\bigg(U_1(t)^2+U_2(t)^2\bigg)\bigg] \eta=0.
    % + \frac{\mathrm{i} (U_1 \rho_1+U_2\rho_2) }{\tanh \frac{kh}{2}}  \frac{\mathrm{d}\eta}{\mathrm{d}t} 
 %   - \frac{k^2( U_1^2 \rho_1 + U_2^2\rho_2) }{\left(\rho_1+\rho_2\right)} \eta + T\frac{kg^\prime}{2} \cos \left(\alpha(t)\right) \eta=0
\end{align}
%where $T=\tanh \left(kh/2\right)$.
The explicit time dependence of the background flow has been intentionally emphasised.
For a tank at rest, meaning $\alpha\!=\!0$ and $U_1\!=\!U_2\!=\!0$, we recover the equation governing interfacial gravity waves with frequency $\omega=\pm \sqrt{(g'k/2)\tanh(kh/2)}$ propagating in a two-layered fluid where each layer has a depth of $h/2$, see \cite{sutherland2010internal}. Hereafter we will restrict our attention to $\tanh \left(kh/2\right)\!\approx\!1$, i.e.\,the waves are much shorter than the depth of the individual layers. The justification behind this choice is motivated by the fact that the wavelengths of KH instabilities arising in the pycnocline are usually much smaller than the pycnocline's depth. Hence, the layer depths do not play any further role in governing the dynamics. 

Substituting \eqref{eq:base_vel_simp_a}--\eqref{eq:base_vel_simp_b} into \eqref{eqn:eta_gen} yields
 \begin{align}
    \eta_{,tt} + \left[ \omega^{2} \left(1-
     \frac{\alpha_f^2}{4} (1-\cos(2\omega_ft))\right) - \omega^{4}\frac{\alpha_f^2}{\omega_f^2} \left( \frac{3}{2} - 2 \cos\left(\omega_ft\right) + \frac{1}{2}\cos\left(2\omega_f t\right)\right)\right]\eta=0.
     \label{eqn:eta_with_all_params}
 \end{align}
% {\color{blue} 
% The forcing oscillation with frequency $\omega$ generates long interfacial wave. In the absence of the forcing $\alpha_f=0$, the interface between two fluid is perturbed by an interfacial wave of frequency $\omega$. } 
Hereafter, $\omega_f$ will represent low-frequency oscillations of the density interface. Physically, this can represent low-order, basin-scale modes \citep[Chapter 2.5]{sutherland2010internal} which are essentially standing waves equivalent to seiches in lakes. Furthermore, although the analogy is tenuous,  $\omega_f$ can also represent the frequency of propagating long interfacial waves; see \cite{inoue2009efficiency}. In any case, the two-layered oscillating tank setup in figure \ref{fig:schematic} is a simplified representation of the realistic scenario, and the key outcome in this context is  the appearance of a small parameter $\epsilon\!=\!\omega_f/\omega\!\ll\!1$. In other words, the frequencies (wavelengths) of the gravity waves at the density interface, which could be rendered unstable, are orders of magnitude greater (smaller) than the frequencies (wavelengths) characterising the mean density interface, i.e.\,the background flow. The mean density interface represents a small amplitude long wave of infinite wavelength, which essentially makes it a straight line and further ensures that the background shear flow is parallel.
%Noting that $\omega_f$ corresponds to low-frequency oscillations and may be regarded as a proxy for the frequency of semi-diurnal internal tides, we immediately arrive at a small parameter $\epsilon=\omega_f/\omega\!\ll\!1$. Furthermore, the condition $\epsilon\!\ll\!1$ is also essential for ensuring parallel shear flow conditions (which is the basis of \S \ref{subsec:bg_flow}), see \cite{inoue2009efficiency}. 
%We re-emphasise here that the objective is to understand how a short interfacial wave of (high) frequency $\omega$ is rendered unstable by the shear induced by a long interfacial wave of (low) frequency $\omega_f$ (whose effect is incorporated by oscillating the tank). This yields a small parameter $\epsilon=\omega_f/\omega$. 

Defining a nondimensional time $\tau\! =\!\omega_f t$, \eqref{eqn:eta_with_all_params} yields
%  We can rescale time  as $\tau = \omega_f t$,
%   \begin{align}
%     \omega_f^2\frac{\mathrm{d}^2\eta}{\mathrm{d}\tau^2} + \left[ \omega^{2} \left(1-
%      \frac{\alpha_f^2}{4} (1-\cos(2\tau))\right) - \omega^{4}\frac{\alpha_f^2}{\omega_f^2} \left( \frac{3}{2} - 2 \cos\left(\tau\right) + \frac{1}{2}\cos\left(2\tau\right)\right)\right]\eta=0.
%  \end{align}
% Let us assume $\epsilon=\omega_f/\omega$. 
% Then the above equation reduces to
 % \begin{align}\label{eqn:eta_with_eps}
 %    \epsilon^2\frac{\mathrm{d}^2\eta}{\mathrm{d}\tau^2} + \left[ \left(1-
 %     \frac{\alpha_f^2}{4} (1-\cos(2\tau))\right) - \frac{\alpha_f^2}{\epsilon^2} \left( \frac{3}{2} - 2 \cos\left(\tau\right) + \frac{1}{2}\cos\left(2\tau\right)\right)\right]\eta=0.
 % \end{align}
% Note that, $\epsilon^2 \ll 1$. Consider $\beta = \alpha_f^2/\epsilon^2$,
 \begin{align}\label{eqn:eta_with_beta}
    \epsilon^2\eta_{,\tau \tau} + \bigg[\underbrace{1-   \frac{3}{2} \beta
    + \beta \left( 2 \cos\left(\tau\right) - \frac{1}{2}\cos\left(2\tau\right)\right) }_{\text{$\mathcal{F}(\tau)$}}-   \frac{\alpha_f^2}{4} \big(1-\cos(2\tau)\big) \bigg]\eta=0,
 \end{align}
 where $\beta\!=\!\alpha_f^2/\epsilon^2$. Comparing $\beta$ with the definition of minimum Richardson number $Ri_{\textrm{min}}$ in \citet[Eq.\,(13)]{inoue2009efficiency}, we observe that $\beta$ resembles the inverse of minimum Richardson number. However, we are cautious that there is no length scale (shear layer thickness) in our problem, hence a direct comparison with \cite{inoue2009efficiency} might be misleading.   Note that $\beta\!\gg\!\alpha_f^2$, thus  $\mathcal{O}(\alpha_f^2)$ terms can be neglected. This transforms \eqref{eqn:eta_with_beta} into the following governing equation of the interface, which  can be classified as a Schr\"{o}dinger-type equation with a periodic potential:
 \begin{align}
     \epsilon^2\eta_{,\tau \tau} + \mathcal{F}(\tau) \eta=0 \quad \mathrm{where} \quad\mathcal{F}(\tau)=\mathcal{F}(\tau+2\pi).
    \label{eqn:eta_simplified_ver}
 \end{align}

It is worthwhile to compare \eqref{eqn:eta_simplified_ver} with the governing ODE obtained by \citet[Eq.\,(3.17)]{kelly1965stability}. Kelly's two-layered flow with imposed layer-wise oscillating shear is \textit{only} applicable to non-Boussinesq flows (shear disappears under the Boussinesq limit, see \citet[Eq.\,(2.3)]{kelly1965stability})  and hence is fundamentally different from our system.  The ODE governing Kelly's system is Mathieu-type and therefore undergoes parametric instability when the forcing frequency $\omega_f$ is twice the natural frequency $\omega$. Such parametric resonance is \textit{not} expected to be the mechanism driving instability in our system since  $\omega$ and $\omega_f$ differ by many orders of magnitude. Furthermore,  previous authors \citep{inoue2009efficiency,lewin2022stratified} do not report any finely tuned frequency ratios for observing the instability. The order separation of the two frequencies gives rise to the coefficient $\epsilon^2$ in the second derivative, thereby rendering \eqref{eqn:eta_simplified_ver} into a  Schr\"{o}dinger-type equation.  We note in passing that \cite{onuki2021simulating} numerically studied the situation for which $\omega\!\sim\!\omega_f$ (hence $\epsilon\!\sim\!1$), and indeed the authors report parametric subharmonic instability. 

\subsubsection{Necessary and sufficient condition for instability}
 \label{subsubsec:CONDIND}
\begin{figure}
    \centering
    \includegraphics[scale=0.35]{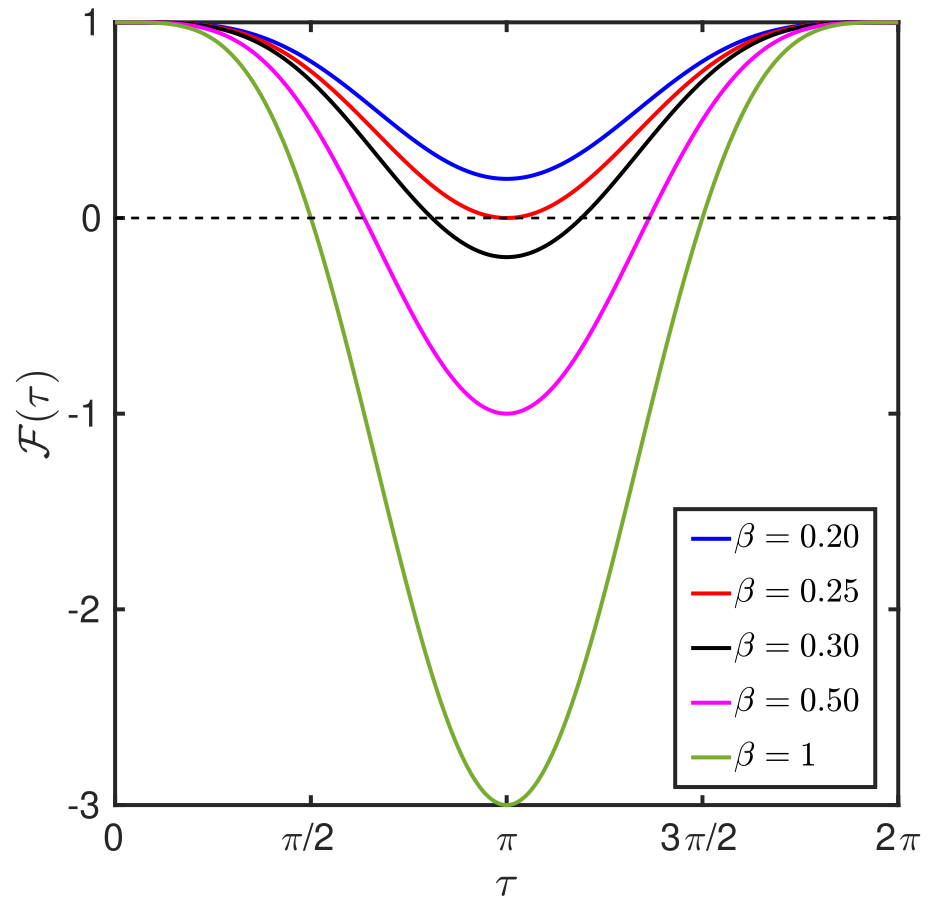}
    \caption{Variation of $\mathcal{F}(\tau)$ with $\tau$ for different $\beta$ values. The flow becomes unstable when $\mathcal{F}(\tau)<0$. For reference, $\tau\!=\!\pi/2$ implies the maximum counter-clockwise excursion (i.e. $\alpha=\alpha_f$) of the two-layered oscillating tank, while $\tau\!=\!\pi$ implies that the tank, through clockwise rotation, has reached the horizontal position.}
    \label{fig:F_vs_tau}
\end{figure}

The nature of the solution to \eqref{eqn:eta_simplified_ver} is  determined by the sign of $\mathcal{F}(\tau)$; instability implies $\mathcal{F}(\tau)\!<\!0$ while the flow is stable for  $\mathcal{F}(\tau)\!>\!0$.
%The solution of \eqref{eqn:eta_simplified_ver} is oscillatory or remains bounded if $\mathcal{F}(\tau)$ is positive and exhibits exponential growth if $\mathcal{F}(\tau)$ is negative. The condition for instability is defined by the inequality
After some algebra, the necessary and sufficient condition for instability is found to be
\begin{align}
    \cos \tau < 1-\frac{1}{\sqrt{\beta}} \implies \beta > \beta_{\mathrm{min}} = \frac{1}{4}.
    \label{eq:stab_cond}
\end{align}
This implies all waves with wavenumbers $k>k_{\mathrm{long-wave-cut-off}}=\omega_f^2/(2g'\alpha_f^2)$ are unstable.
Figure~\ref{fig:F_vs_tau} represents the variation of $\mathcal{F}(\tau)$ with $\tau$ for various $\beta$ values. Initially, the flow is stable for all finite $\beta$ values. For $\beta\!>\!0.25$, the solution `tunnels' into the unstable region after crossing the (first) turning point $\tau=\tau_c$, which signifies $\mathcal{F}(\tau)$ changing sign from positive to negative.
From \eqref{eq:stab_cond}, $\tau_c$ can be straightforwardly determined:
\begin{align}
    \tau_c = \cos^{-1} \bigg(1-\frac{1}{\sqrt{\beta}}\bigg).
    \label{eq:tau_c}
\end{align}
Noting that $\mathcal{F}(\tau)$ has a period of $2\pi$ and a mirror symmetry about $\tau=\pi$, it changes sign (for $\beta\!>\!0.25$)  twice within its period -- first at $\tau=\tau_c$ and then at $\tau=2\pi-\tau_c$. The stability condition provides an upper bound for the time of occurrence of the instability -- if the flow is unstable, then instability will set in before $\tau=\pi$.  The second turning point at $\tau=2\pi-\tau_c$ signifies stabilisation, which may have little effect for higher $\beta$ values that allow sustained growth for a longer time interval ($2\pi-2\tau_c$). This is because $\tau_c$ reduces as $\beta$ increases. Hence, the role of  $\beta$ is analogous to that of the inverse of the Richardson number. 

In Appendix \ref{appA}, we discuss the generic situation in which the two fluid layers have unequal depths: $h_1$ and $h_2\!=\!h-h_1$.  In this scenario, the necessary and sufficient condition for instability is given by (see \eqref{eq:gen_inst_cond})
\begin{align}
    \beta > \beta_{\mathrm{min}}= \frac{1}{8\left[r^2+(1-r)^2\right]},
    \label{eq:beta_unequal_depth}
\end{align} 
where $r=h_1/h$. Since $r\in(0,1)$, we must have $1/8\!<\!\beta_{\mathrm{min}} \!\leq \! 1/4$.  The upper bound for the time of occurrence of the instability is independent of $r$ and occurs at $\tau\!=\!\pi$.
In the oceanographic context, the typical depth of the pycnocline ($h_1$) is much less than the total ocean depth ($h$), i.e. $r\!\ll\!1$, which leads to the instability condition  $\beta\!>\!1/8$.

 \subsubsection{Predictions  from the steady background flow assumption}
\label{subsubsec:normal_mode_pred}

 Undertaking a conventional, steady-state-based linear stability analysis for an inherently unsteady background flow could appear simplistic. However, such an analysis has a key advantage -- the simplifications can result in an exact solution of \eqref{eqn:eta_simplified_ver}, leading to a closed-form expression for the growth rate. Furthermore, if the growth is exponential (verified \textit{aposteriori}), then the time-scale of growth is much faster than the slow time dependence of the background flow, implying that the steady-state assumption is sensible.
 Here we intend to explore the predictions obtained from steady-state-based analyses and the extent of errors incurred.

One way of removing the explicit time dependence of the background flow is time-averaging. 
By averaging \eqref{eq:base_vel_simp_a}--\eqref{eq:base_vel_simp_b} over its period, we obtain 
 \begin{equation}
    \overline{U_1}=-\overline{U_2}=\frac{g^\prime}{2} \frac{\alpha_f}{\omega_f}, 
\end{equation}
which when substituted into \eqref{eqn:eta_gen} converts  \eqref{eqn:eta_simplified_ver} into a second-order \textit{constant} coefficient ODE with 
 $\mathcal{F}=1-\beta$. This yields the instability condition $\beta\!>\!1$, which implies normal mode instabilities with a growth rate of $\sigma_{\textrm{normal-mode}}\!=\!\sqrt{\beta-1}/\epsilon$. 
% \begin{equation}
% \beta>1 \implies k > \frac{h}{2a_f^2}.
% \label{eq:beta_1_wrong}
% \end{equation}
Comparison with \eqref{eq:stab_cond} shows that the above instability condition is erroneous.
%This flow is unstable to normal mode instabilities with a growth rate of $\sigma_{\textrm{normal-mode}}\!=\!\sqrt{\beta-1}/\epsilon$. A comparison between \eqref{eq:stab_cond}  and \eqref{eq:beta_1_wrong} reveals that the latter is erroneous.
%; the minimum wavenumber experiencing instabilities is wrongly predicted to be  $4$ times larger than the result obtained from the unsteady analysis, see \eqref{eq:lambda_max}. 

%Furthermore, the steady-state-based stability analysis predicts the minimum wavenumber experiencing instabilities to be  $4$ times larger than the one predicted from the unsteady analysis, see \eqref{eq:lambda_max}.  
%With regards to the case study in \S \ref{subsubsec:case_study}, steady-state-based stability analysis would predict instabilities for all wavelengths smaller than $0.34$m (where unsteady analysis predicts $1.35$m).

Next, we undertake another steady-state-based approach where the maximum background velocity of each layer given in \eqref{eq:base_vel_simp_a}--\eqref{eq:base_vel_simp_b} is considered:
\begin{equation}
    U_{1,\textrm{max}}=-U_{2,\textrm{max}}=g^\prime\frac{\alpha_f}{\omega_f}. 
    \label{eq:max_vel_steady}
\end{equation}
The above background velocities are substituted in \eqref{eqn:eta_gen} yielding  $\mathcal{F}=1-4\beta$ in the governing equation \eqref{eqn:eta_simplified_ver}. This flow becomes unstable to normal mode instabilities which has a growth rate of 
\begin{equation}
    \sigma_{\textrm{normal-mode}}\!=\!\frac{\sqrt{4\beta-1}}{\epsilon}.
    \label{eq:NMmode}
\end{equation}
 The resulting instability condition $\beta\!>\!1/4$
exactly matches the unsteady analysis in \eqref{eq:stab_cond}.
%Note that \eqref{eq:max_vel_steady}, when written in terms of the long wave's amplitude and phase speed ($c_f$) would yield
%\begin{equation*}
%    U_{1,\textrm{max}}=-U_{2,\textrm{max}}=\frac{4c_f a_f}{H}
%\end{equation*}
%for the case where the interface is located at a depth of $H/2$. 
To see how this growth rate compares with the classical KH instability of two discontinuous layers, we write the well-known expression for its dimensional growth rate \citep{drazin2004,guha2019predicting}:
\begin{equation}
\widetilde{\sigma}_{\textrm{KH}}\!=\!\sqrt{U^2k^2-\frac{g'k}{2}},
    \label{eq:classicalKH}
\end{equation}
where the upper (lower) layer of density $\rho_1\,(\rho_2)$ is moving with a constant velocity $U\,(-U)$. Subsituting $U\!=\!U_{1,\textrm{max}}$ from  \eqref{eq:max_vel_steady} and nondimensionalising \eqref{eq:classicalKH} with $\omega_f$, it is straightforward to verify that the nondimensional growth rate ${\sigma}_{\textrm{KH}}\!=\!{\widetilde{\sigma}_{\textrm{KH}}}/{\omega_f}\!=\!\sigma_{\textrm{normal-mode}}$ in \eqref{eq:NMmode}. 

In summary, the steady-state-based linear stability analysis that considers the maximum  (instead of time-average) background velocity in each layer provides the same instability condition as the time-dependent problem. The normal mode growth rate exactly matches the classical KH instability. However, such an analysis is unable to capture key features of the actual system, which is inherently unsteady. For example, the steady-state-based analysis cannot predict the time of onset of the instability. 
%For example,  the steady-state-based analysis predicts exponential growth from time $\tau\!=\!0+$, while the unsteady analysis in \S \ref{subsubsec:CONDIND} shows that the instability sets in after $\tau\!=\!\tau_c$. 
Since $\tau$ is a slow time, a delayed onset of instability is a salient feature of this system.  How well the growth rate predicted from the steady-state-based analysis compares with the unsteady analysis also needs to be examined.

\section{Linear stability analysis}
\label{sec:linstab}
\subsection{Floquet analysis, numerical integration, and normal mode predictions}
\label{subsec:Floquet}
Equation \eqref{eqn:eta_simplified_ver} can be written as a 2D dynamical system
\begin{equation}
    \begin{bmatrix}
         \eta  \\
         v 
     \end{bmatrix}_{,\tau} =\underbrace{\begin{bmatrix}
       0  &  1 \\
       -{\epsilon^{-2}}\mathcal{F}(\tau) & 0
    \end{bmatrix}}_{\text{$\mathcal{L}(\tau)$}} \begin{bmatrix}
         \eta  \\
         v
     \end{bmatrix},  
     \label{eq:2Dsysm}
     \end{equation}
where the linear operator $\mathcal{L}(\tau)$ has a fundamental period of 
 $T\!=\!2\pi$. The fundamental solution matrix $\Phi (\tau)$ satisfies
\begin{equation*}
     \Phi_{,\tau}(\tau) = \mathcal{L}(\tau)\Phi(\tau) \quad \text{where} \quad \Phi(0)=I.
\end{equation*}

Floquet analysis is the standard technique for analysing the stability of linear periodic systems. If the analysis finds that the energy of the disturbance at the end of one period is larger than the energy at the outset, the system is deemed unstable. In this regard, the monodromy matrix $\Phi(T)$ (which is numerically computed in general) is first determined and its eigenvalues,   known as 
 Floquet multipliers $\mu$, are sought. The Floquet exponents $\Omega$, signifying the complex temporal growth rate, are related to the Floquet multipliers via the relation $\mu\!=\!\exp{(\Omega T)}$. The system is deemed unstable if there exists at least one eigenvalue satisfying $|\mu|\!>\!1$ ($\mathrm{Re}(\Omega)\!>\!0$). Likewise, the system is stable if all eigenvalues satisfy $|\mu|\!<\!1$ ($\mathrm{Re}(\Omega)\!<\!0$). When  all eigenvalues satisfy $|\mu|\!=\!1$ ($\mathrm{Re}(\Omega)\!=\!0$), the system oscillates between finite bounds $\forall \tau$ \citep{richards2002advanced}. In this case, the solution is neutrally stable and can yield quasiperiodic behaviour in time \citep{kovacic2018mathieu}.

We compute  Floquet exponents and the Floquet multipliers using Matlab corresponding to different $\beta$ values in Table \ref{tab:1}. We keep $\alpha_f$ fixed at $0.05$ (which is equivalent to $2.86$\textdegree) in the entire paper.  The system is bounded for $\beta\!=\!0.1$ and $0.2$, and unstable for $0.27$, $0.3$ and $0.35$, which are the expected outcomes as per the condition in \eqref{eq:stab_cond}. Furthermore, we also observe that when $\beta\!>\!0.25$, the higher the $\beta$ value, the higher the maximum growth rate $\mathrm{Re}(\Omega_\textrm{max})$. We restrict our Floquet analysis up to $\beta\!=\!0.35$ since beyond this value, the numerical solution becomes unreliable owing to the resulting extremely large and extremely small Floquet multipliers.

%Two initial conditions are required for solving \eqref{eqn:eta_simplified_ver}. In this regard, we choose the following legitimate initial conditions 

\begin{table}
    \centering
    \begin{tabular}{ccccccccccccccccccccc}
        $\beta$ & & & &  $\mu_{\textrm{1}}$ & & & & $\mu_{\textrm{2}}$ & & & & $|\mu_{\textrm{max}}|$ & & & & $\mathrm{Re}(\Omega_\textrm{max})$ \\ 
        \hline
         $0.1$ & & & & $0.3564 - 0.9343$i & & & & $0.3564 + 0.9343$i & & & & $1$& & & & $0$\\
                  $0.2$ & & & & $-0.1704 - 0.9854$i & & & & $-0.1704 + 0.9854$i & & & & $1$& & & & $0$\\
            $0.27$ & & & & $0.3108$ & & & & $3.2176$ & & & & $3.2176$ & & & & $0.1860$\\      
         $0.3$ & & & & $0.0120$ & & & & $83.1471$ & & & & $83.1471$ & & & & $0.7036$\\
         $0.35$ & & & &   $6.9617\mathrm{E}\!-\!04$ & & & &  $1.4364\mathrm{E}\!+\!03$ & & & & $1.4364\mathrm{E}\!+\!03$ & & & & $1.1570$
    \end{tabular}
    \caption{Floquet analysis of the system \eqref{eq:2Dsysm}. }
    \label{tab:1}
\end{table}

\begin{figure}
    \centering
    \begin{subfigure}{0.47\textwidth}
        \includegraphics[scale=0.22]{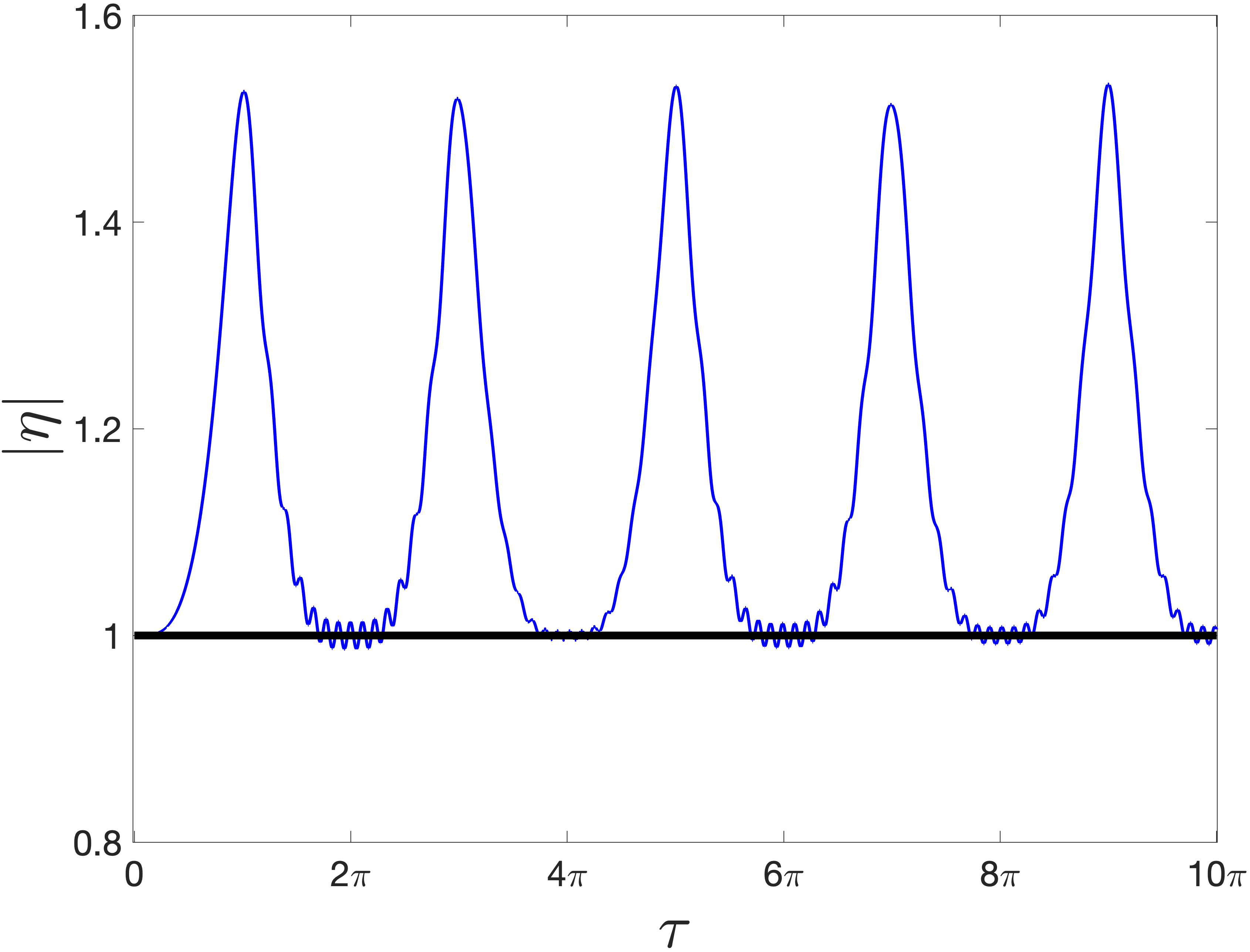}
    \caption{$\beta=0.2$}
    \label{fig:beta_20}
\end{subfigure}
%\hfill
\begin{subfigure}{0.47\textwidth}
        \includegraphics[scale=0.22]{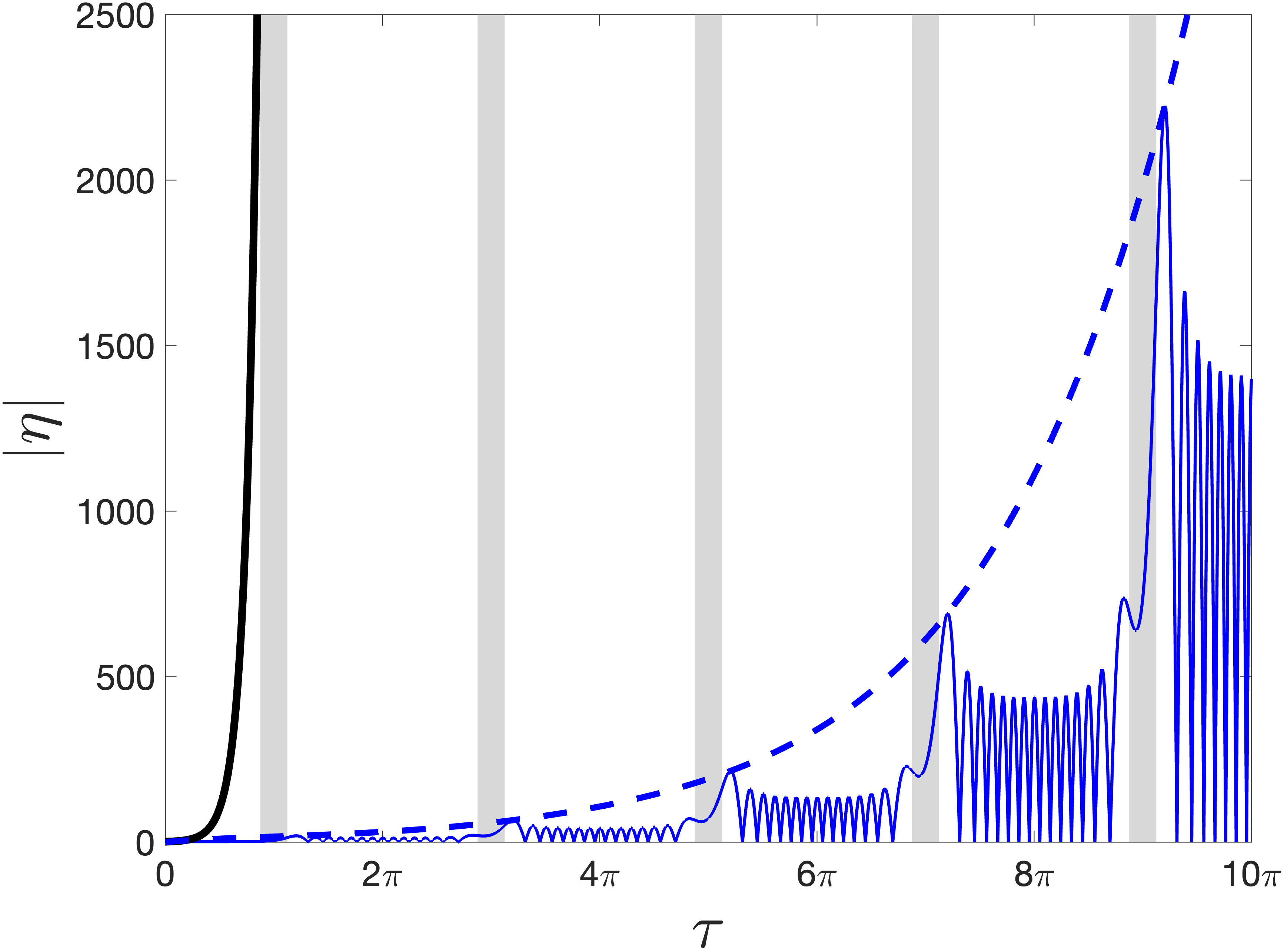}
    \caption{$\beta=0.27$}
    \label{fig:beta_27}
\end{subfigure}
\begin{subfigure}{0.47\textwidth}
    \includegraphics[scale=0.223]{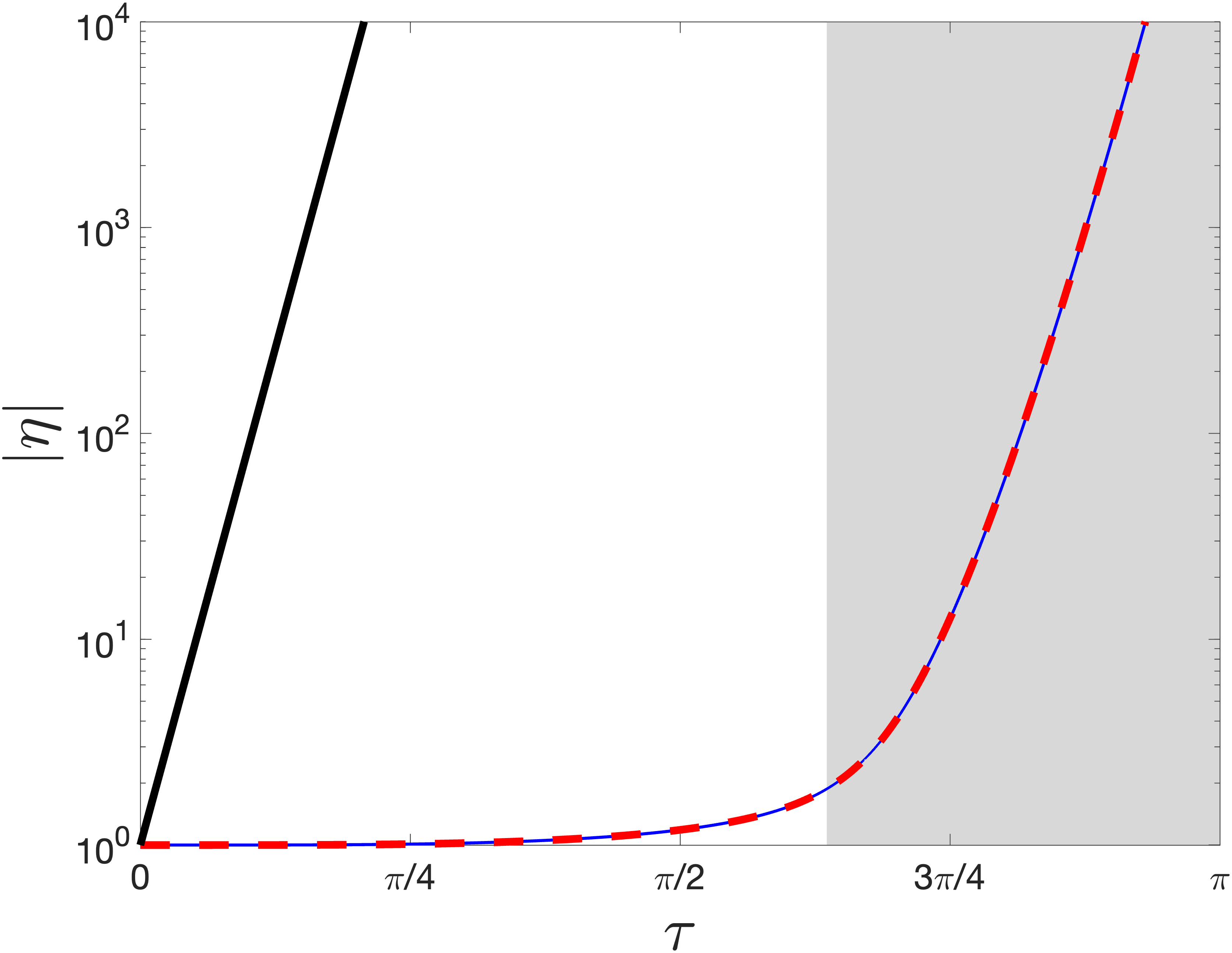}
    \caption{$\beta=0.5$}
     \label{fig:beta_p5}
\end{subfigure}
\hspace{1.5mm}
\begin{subfigure}{0.47\textwidth}
    \includegraphics[scale=0.225]{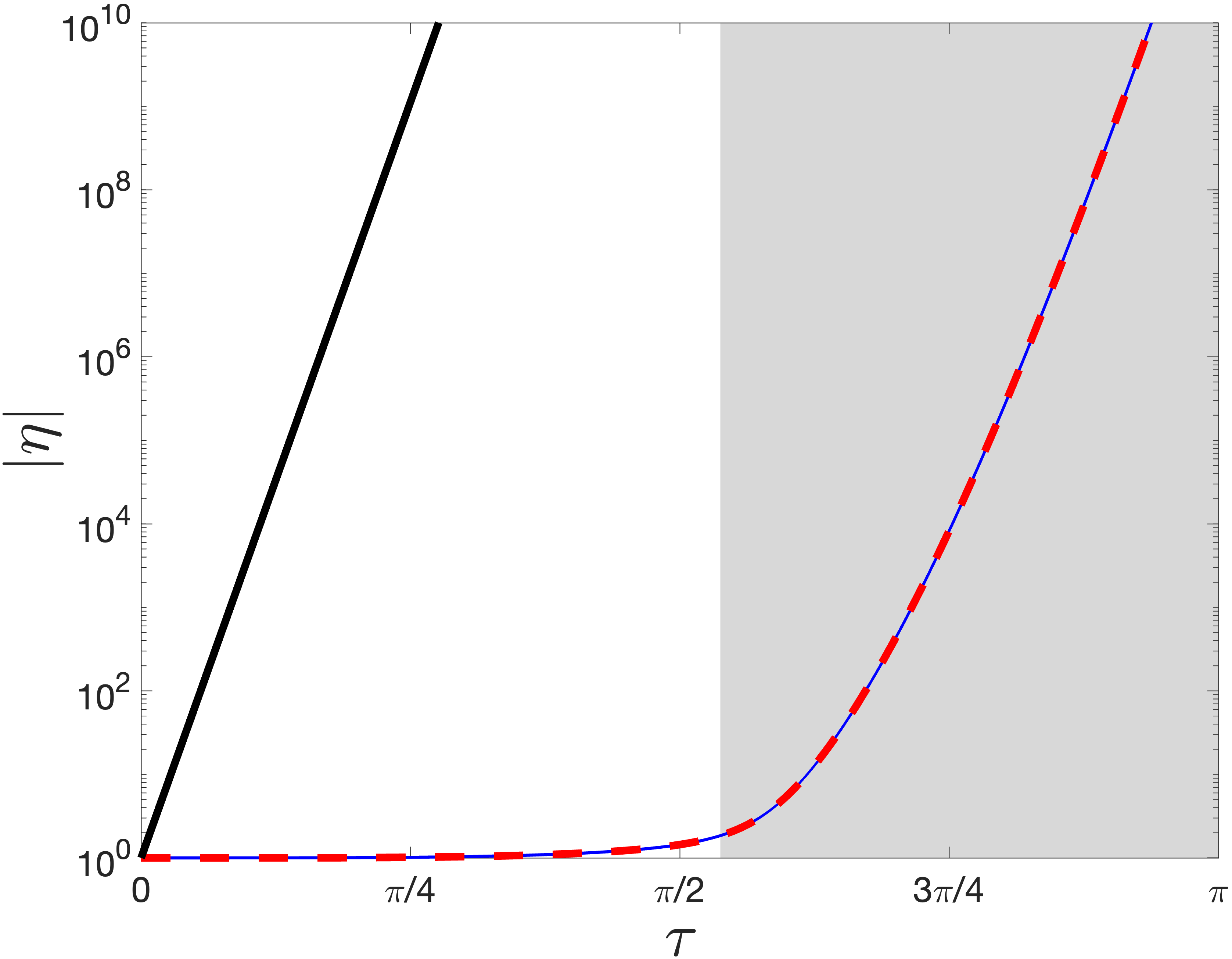}
    \caption{$\beta=0.8$}
   \label{fig:beta_80}
\end{subfigure}
\caption{Plots of $|\eta|$ vs $\tau$ for different $\beta$ values. Unstable regions predicted from \eqref{eq:tau_c} are shaded in grey. Line colours are as follows: black - normal mode solution (based on steady background flow \eqref{eq:max_vel_steady}), blue - numerical solution of \eqref{eqn:eta_simplified_ver}, and dashed-red - MAF solution \eqref{eq:MAF_final}. The dashed-blue line in (b) represents the amplitude envelope.} 
\label{fig:comaprisons_all}
\end{figure}

To understand how the amplitude evolves in time, and to explore a bigger range of $\beta$ values, we numerically solve \eqref{eqn:eta_simplified_ver} using Matlab's inbuilt ODE45 solver (which was also used to construct the monodromy matrix while performing the Floquet analysis). For the two initial conditions, we take
\begin{align}\label{eqn:initial_cond}
   \eta(0)=1 \quad \text{and} \quad  \eta_{,\tau}(0) =\frac{\mathrm{i}}{\epsilon},
\end{align}
which satisfy the governing equation
\begin{align}
    \epsilon^2 \eta_{,\tau \tau} + \eta=0.
\end{align}
The above equation, which can be obtained by taking the limit $\tau \rightarrow 0$ of  \eqref{eqn:eta_simplified_ver}, physically represents propagating deep water interfacial waves inside the two-layered horizontal tank at rest (i.e.\,non-oscillating).
%\color{blue}In the limit $\tau \rightarrow 0$, the above equation reduces to 

Figure \ref{fig:beta_20} shows that the system undergoes bounded oscillations for $\beta\!=\!0.2$, which cannot be predicted from the normal mode theory. %while figure~\ref{fig:beta_27}, corresponding to $\beta\!=\!0.27$, shows instability.
As already noted in \S \ref{subsec:lin_per},  systems with $\beta\!>\!0.25$ start becoming unstable when $\tau\!>\!\tau_c$. Within the $n$-th period, the system will grow when $2n\pi+\tau_c \!<\!\tau\!<\!2(n+1)\pi-\tau_c$,
i.e.\,the interval when $\mathcal{F}(\tau)\!<\!0$ is satisfied. This fact is evident in figure~\ref{fig:beta_27} which corresponds to $\beta\!=\!0.27$. Furthermore, the plot also reveals the amplification of energy between consecutive periods. Figures \ref{fig:beta_p5} and \ref{fig:beta_80}  respectively correspond to $\beta\!=\!0.5$ and $\beta\!=\!0.8$, and reveal a much larger growth rate, as is expected from the trend observed in table \ref{tab:1}. 

Finally, we compare the growth rate $\sigma_{\textrm{numerical}}$ obtained from the numerical solutions (calculated when $\tau\!>\!\tau_c$) with the normal mode  predictions in \eqref{eq:NMmode}:

\vspace{3mm}

\begin{enumerate}
   \item $\beta\!=\!0.27$: \hspace{1.8mm} $\sigma_{\textrm{normal-mode}}=2.94$  \hspace{4.1mm} $\sigma_{\textrm{numerical}}=0.18$ (amplitude envelope   growth rate)
    \item $\beta\!=\!0.5$: \hspace{4mm} $\sigma_{\textrm{normal-mode}}=14.14$  \hspace{2.3mm} $\sigma_{\textrm{numerical}}=14.12$
        \item $\beta\!=\!0.8$: \hspace{4mm} $\sigma_{\textrm{normal-mode}}=26.53$  \hspace{2.3mm} $\sigma_{\textrm{numerical}}=26.53$
\end{enumerate}

\vspace{3mm}

 \noindent As evident from above data and also from figures \ref{fig:beta_p5} and \ref{fig:beta_80}, the normal mode growth rate predictions from \eqref{eq:NMmode} provide an excellent agreement with the numerical solution, and this agreement improves even further as $\beta$ increases. Since the normal mode growth rate exactly matches with that of the classical KH instability, this conclusively proves that the instability is of KH type. However, for the $\beta\!=\!0.27$ (which is a $\beta$ value slightly greater than the instability condition) case, the normal mode theory fails to predict the growth rate of the amplitude envelope, as well as the intricate temporal dynamics of the amplitude evolution, which is clear from figure \ref{fig:beta_27}. 
 %The only aspect which is well predicted by the normal mode theory is the \textit{maximum} growth rate of the amplitude envelope: $\sigma_{\textrm{normal-mode}}=2.94$  while  $\sigma_{\textrm{numerical}}=2.96$.
 
 We also emphasise here that $\beta$, by definition, is proportional to  $\omega^2\!=\!g'k/2$. The growth rate trend reveals the classical `ultraviolet catastrophe', i.e.\,shorter waves are more unstable, which is an artefact of the absence of a lengthscale (i.e.\,finite thickness of the interface) in the problem.
 %and is observed in similar setups with interfacial discontinuity, e.g.\,the classical Rayleigh-Taylor and  Kelvin-Helmholtz instabilities \citep{drazin2004}.   

 Finally, we compare our theoretical predictions with the DNS results of \cite{lewin2022stratified}. In this regard, we choose the case `NM72D', where normal mode perturbations have been used to seed the flow. Note that the authors use smooth profiles, hence, their parameters need to be suitably adjusted to fit our discontinuous setup. Their parameters are as follows: $\alpha_f\!=\!7.28^\circ\!=\!0.127$ radians, $\lambda\!=\!14.28$ implying $k\!=\!0.44$, half shear layer thickness $\delta\!=\!1$, and $\omega_f/N_b\!=\!0.072$, where $N_b$ denotes background buoyancy frequency. Noting that $g'\!=\! N_b^2\delta$, we can state $\epsilon^2$ in their variables:
\begin{equation*}
    \epsilon^2=\frac{\omega_f^2}{\omega^2}=\frac{2}{k\delta}\bigg(\frac{\omega_f}{N_b}\bigg)^2=0.024.
\end{equation*}
This yields
\begin{equation*}
   \beta=\frac{\alpha_f^2}{\epsilon^2}=0.67>\frac{1}{4},
\end{equation*}
and thus the setup for NM72D satisfies our instability condition \eqref{eq:stab_cond}. Indeed, the authors observe KH instability for the case NM72D. Furthermore, the onset of instability predicted from \eqref{eq:tau_c} is $\tau_c\!=\!0.57\pi$, which excellently matches the corresponding DNS result $\tau_c\!=\!0.54\pi$  (see Acknowledgement). However, we provide a note of caution that while certain instability characteristics (e.g.\,$\tau_c$) can have a direct quantitative agreement between discontinuous and smooth profiles, this does not guarantee a correspondence between other instability characteristics. For example, growth rates obtained from discontinuous profiles are unreliable. Moreover, discontinuous profiles cannot predict the band of unstable wavenumbers or the most unstable wavenumber. A proper evaluation of these parameters demands consideration of a finite-thickness shear layer.

 \begin{subsection}{Asymptotic solution: the Modified Airy Function (MAF) method}
 \label{subsec:MAF}
Here we aim to construct an approximate solution to the amplitude evolution. A standard approach in this regard would be to employ the well-known singular perturbation technique -- the Wentzel–Kramers–Brillouin (WKB) method \citep{bender2013advanced}. 
% The well known  Wentzel–Kramers–Brillouin (WKB) theory -- a singular perturbation technique that provides global approximation to the solution of Schr\"{o}dinger-type equations \citep{bender2013advanced}, has been used for solving \eqref{eqn:eta_simplified_ver}.
% %Here we apply the WKB method to obtain an asymptotic solution of~\eqref{eqn:eta_simplified_ver}.
In the WKB method, the asymptotic series in the small parameter $\epsilon$ is written as follows
%Thus, we assume the solution in the form of an asymptotic series  as follows,
 \begin{align}\label{eqn:WKB_gen_expansion}
  \eta(\tau) \sim \exp{\left(\frac{1}{\epsilon} \sum\limits_{n=0}^{\infty} \epsilon^n S_n(\tau)\right)},
  %\qquad \text{in the limit}~~ \epsilon \rightarrow 0.
\end{align}
where $S_n(\tau)$ denotes the complex phase. Substitution of  the above expansion in \eqref{eqn:eta_simplified_ver} yields the following first order expressions for $\eta$ (valid for $0\!\leq\!\tau\!<\! 2\pi-\tau_c$, i.e. up to the second turning point:
{
\begin{equation}
 \eta(\tau) = \left\{
 \begin{array}{lll}
 C_1|\mathcal{F}(\tau)|^{-\frac{1}{4}}   \exp\left(\frac{\mathrm{i}}{\epsilon} \int_{0}^{\tau} \sqrt{|\mathcal{F}(\tilde{\tau})|}\,\mathrm{d}\tilde{\tau}\right) &\text{for} & 0\!\leq\!\tau\!<\!\tau_c,\\ 
 C_2|\mathcal{F}(\tau)|^{-\frac{1}{4}}   \exp\left(\frac{1}{\epsilon} \int_{\tau_c}^\tau \sqrt{|\mathcal{F}(\tilde{\tau})|}\,\mathrm{d}\tilde{\tau}\right) \\ \,\,\,\,\,\,\,\,\,\,\,\,+C_3|\mathcal{F}(\tau)|^{-\frac{1}{4}}   \exp\left(-\frac{1}{\epsilon} \int_{\tau_c}^\tau \sqrt{|\mathcal{F}(\tilde{\tau})|}\,\mathrm{d}\tilde{\tau}\right) & \text{for} &\tau_c\!<\!\tau\!<\!2\pi-\tau_c.
%  (\mathcal{F}(\tau))^{-\frac{1}{4}}   \cos\left(\frac{1}{\epsilon} \int_{\tau_c}^{\tau} \sqrt{\mathcal{F}(\tilde{\tau})}\,\mathrm{d}\tilde{\tau} +\frac{\pi}{4}\right) & \text{for} & \tau > \tau_c.
 \end{array}
 \right. 
 \label{eq:WKB_sol}
\end{equation}
% \begin{equation}
%  \eta(\tau) =  \left\{
%  \begin{array}{lll}
% C_1 |\mathcal{F}(\tau)|^{-\frac{1}{4}}   \exp\left(\frac{\mathrm{i}}{\epsilon} \int_{0}^{\tau} \sqrt{|\mathcal{F}(\tilde{\tau})|}\,\mathrm{d}\tilde{\tau}\right) & \text{for} & 0\!\leq\!\tau\!<\!\tau_c,\\ 
% C_2 |\mathcal{F}(\tau)|^{-\frac{1}{4}}   \exp\left(\frac{1}{\epsilon} \int_{\tau_c}^\tau \sqrt{|\mathcal{F}(\tilde{\tau})|}\,\mathrm{d}\tilde{\tau}\right) & \text{for} &\tau_c\!<\!\tau\!<\!2\pi-\tau_c.
% %  (\mathcal{F}(\tau))^{-\frac{1}{4}}   \cos\left(\frac{1}{\epsilon} \int_{\tau_c}^{\tau} \sqrt{\mathcal{F}(\tilde{\tau})}\,\mathrm{d}\tilde{\tau} +\frac{\pi}{4}\right) & \text{for} & \tau > \tau_c.
%  \end{array}
%  \right. 
%  \label{eq:WKB_sol}
% \end{equation}
%while noting the fact that the above solution is not defined at the critical point $\tau=\tau_c$. 
 The solution exhibits an oscillatory behaviour when $\tau\!<\!\tau_c$ and grows monotonically (the decaying solution would rapidly become insignificant)  when $\tau_c\!<\!\tau\!<\!2\pi-\tau_c$. Note that  $C_1$ can be straightforwardly determined from \eqref{eqn:initial_cond} and yields $C_1\!=\!1$. 
To find $C_2$ and $C_3$, turning point analysis at $\tau\!=\!\tau_c$ is required, which employs Airy functions to match the solutions on the two sides of the turning point. Instead of the standard WKB turning point analysis,  we employ the Modified Airy Function (MAF) method, which can be regarded as an improved variant of WKB. The key aspect of MAF is its  recognition of the fact that the Airy functions
provide a uniformly valid composite approximation both near and far from a turning point.}
MAF was first introduced by \cite{langer1935asymptotic}; for a detailed discussion on this technique, also see \cite{bender2013advanced,MAF_monograph,wine2025modified}. 
Here we employ MAF to analyse both turning points, and hence obtain a uniformly valid solution for the entire period of $2\pi$. In MAF, the amplitude $\eta(\tau)$ is expressed as

  ~%    In this subsection we will discuss the solution of the initial value problem \eqref{eqn:eta_simplified_ver} using the modified airy function (MAF) method. This method yields an approximate analytic solution to the Schr\"{o}dinger-type wave equation. The MAF solution is valid even at the turning points where the WKB solution fails. 
%  The method is detailed in \cite{MAF_monograph}.
\begin{align}\label{eq:maf_gen}
    \eta(\tau) =\left\{ 
    \begin{aligned}
   & F_1(\tau) Ai\left(\zeta_1(\tau)\right) + G_1(\tau) Bi\left(\zeta_1(\tau)\right) \quad \mbox{for} \quad 0\leq\tau\leq \tau_p, \\
    &F_2(\tau) Ai\left(\zeta_2(\tau)\right) + G_2(\tau) Bi\left(\zeta_2(\tau)\right) \quad \mbox{for} \quad  \tau_p\leq\tau\leq2\pi,
    \end{aligned}
    \right. 
\end{align}
where
$\tau_p\!=\!\pi$ is the location of absolute minima of the periodic potential $\mathcal{F}(\tau)$, and
$Ai(\zeta)$ and $Bi(\zeta)$, respectively, the modified Airy functions of the first and second kind, satisfy
\begin{align*}
   Ai_{,\zeta \zeta} - \zeta Ai(\zeta)=0, \quad \mbox{and} \quad  Bi_{,\zeta \zeta} - \zeta Bi(\zeta)=0.
\end{align*}
We will proceed by focusing on one of the terms (say,  $F_1(\tau) Ai\left(\zeta_1(\tau)\right)$) in \eqref{eq:maf_gen},  as the other terms follow similar algebra, and they can be treated independently.
Substituting $F_1(\tau) Ai\left(\zeta_1(\tau)\right)$ into the governing equation \eqref{eqn:eta_simplified_ver} and using the Airy function equations we obtain 
\begin{align*}
    F_{1,\tau\tau} Ai(\zeta) +2 F_{1,\tau} Ai_{,\zeta_1} \zeta_{1,\tau} + F_1 Ai_{,\zeta_1} \zeta_{1,\tau \tau}+ F_1\, Ai(\zeta_1) \left[ \epsilon^2 \zeta_1(\tau) \zeta_{1,\tau}^2 +  \mathcal{F}(\tau) \right]=0. 
 % G_{,\tau\tau} Bi(\zeta) +2 G_{,\tau} Bi_{,\zeta} \zeta_{,\tau} + G Bi_{,\zeta} \zeta_{,\tau \tau}+ G\, Bi(\zeta) \left[ \epsilon^2 \zeta(\tau) \zeta_{,\tau}^2 +  \mathcal{F}(\tau) \right]=0.   
\end{align*}
To eliminate the terms in square brackets,  $\zeta_1(\tau)$ is chosen to satisfy
\begin{align}
    {\epsilon^2} \zeta_1(\tau) \left[\zeta_{1,\tau}(\tau)\right]^2 + {\mathcal{F}(\tau)}=0,
\end{align}
which gives 
\begin{align}
      \zeta_1(\tau) =\left\{ 
    \begin{aligned} -&\left(\frac{3}{2\epsilon} \int_\tau^{\tau_c} \sqrt{\mathcal{F}(\tilde{\tau})} \mathrm{d}\tilde{\tau} \right)^{2/3}  \quad \text{for} \quad 0 \leq \tau \leq \tau_c,\\
    & \left(\frac{3}{2\epsilon} \int_{\tau_c}^{\tau} \sqrt{-\mathcal{F}(\tilde{\tau})} \mathrm{d}\tilde{\tau} \right)^{2/3}  \quad \text{for} \quad \tau_c \leq \tau \leq\tau_p.
    \end{aligned}
    \right. 
\end{align}
$F_1(\tau)$ is obtained by assuming $F_{1,\tau \tau} \approx 0$. Thus, $F_1(\tau)$ satisfies
\begin{align*}
    2 F_{1,\tau} Ai_{,\zeta_1} \zeta_{,\tau} + F_1 Ai_{,\zeta_1} \zeta_{1,\tau \tau}=0,
\end{align*}
which leads to
\begin{align}
    F_1(\tau)= \frac{D_1}{\sqrt{\zeta_{1,\tau}(\tau)}}.
\end{align}
Similarly, $G_1(\tau)$, $F_2(\tau)$, and $G_2(\tau)$ can be expressed as
\begin{align}
    G_1(\tau)= \frac{D_2}{\sqrt{\zeta_{1,\tau}(\tau)}}, \quad F_2(\tau)= \frac{D_3}{\sqrt{\zeta_{2,\tau}(\tau)}}, \quad \mbox{and} \quad G_2(\tau)= \frac{D_4}{\sqrt{\zeta_{2,\tau}(\tau)}},
\end{align}
where \begin{align}
         \zeta_2(\tau) =\left\{ 
    \begin{aligned} &\left(\frac{3}{2\epsilon} \int_\tau^{2\pi-\tau_c} \sqrt{-\mathcal{F}(\tilde{\tau})} \mathrm{d}\tilde{\tau} \right)^{2/3}  \quad  \text{for} \quad \tau_p\leq\tau \leq 2\pi-\tau_c,\\
   - & \left(\frac{3}{2\epsilon} \int_{2\pi-\tau_c}^{\tau} \sqrt{\mathcal{F}(\tilde{\tau})} \mathrm{d}\tilde{\tau} \right)^{2/3}  \quad \hspace{4mm} \text{for}  \quad   2\pi-\tau_c \leq \tau \leq 2\pi.
    \end{aligned}
    \right. 
\end{align}
The arbitrary constants $D_{1},\,D_2,\,D_3$, and $D_4$  will be determined by using the initial conditions \eqref{eqn:initial_cond}  and the continuity of $\eta(\tau)$ and $\eta_{,\tau}(\tau)$ at $\tau_p=\pi$.
% $D_1$ and $D_2$ are arbitrary constants, which can be determined from the initial conditions \eqref{eqn:initial_cond}, and are given below:
The initial conditions \eqref{eqn:initial_cond} give:
\begin{align}
    \begin{bmatrix}
        D_1 \\[6pt]
        D_2
    \end{bmatrix} = \pi  \sqrt{\zeta_{,\tau}(0)} \begin{bmatrix}
        Bi_{,\tau}\left(\zeta_1(0)\right) \\[6pt]
       - Ai_{,\tau}\left(\zeta_1(0)\right)
    \end{bmatrix} - \frac{\pi}{\sqrt{\zeta_{,\tau}(0)}} \left( \frac{\mathrm{i}}{\epsilon}+  \frac{\zeta_{,\tau \tau}(0)}{2\zeta_{,\tau}(0)} \right) \begin{bmatrix}
        Bi\left(\zeta_1(0)\right) \\[6pt]
       - Ai\left(\zeta_1(0)\right)
    \end{bmatrix}. 
\end{align}
The continuity of $\eta(\tau)$ and $\eta_{,\tau}(\tau)$ at $\tau_p=\pi$  yields the following system of linear equations with unknowns $D_3$ and $D_4$:
\begin{align}
    \begin{bmatrix}
        A_3(\tau_p) & B_3(\tau_p)\\[4pt]
        A_{3,\tau}(\tau_p) & B_{3,\tau}(\tau_p) 
    \end{bmatrix} 
    \begin{bmatrix}
        D_3 \\[4pt]D_4
    \end{bmatrix} =  \begin{bmatrix}
        A_2(\tau_p) & B_2(\tau_p)\\[4pt]
        A_{2,\tau}(\tau_p) & B_{2,\tau}(\tau_p) 
    \end{bmatrix} 
    \begin{bmatrix}
        D_1 \\[4pt] D_2
    \end{bmatrix},
\end{align}
% \begin{align}
%     \begin{bmatrix}
%         A_2(\tau_p) & B_2(\tau_p)\\[4pt]
%         A_{2,\tau}(\tau_p) & B_{2,\tau}(\tau_p) 
%     \end{bmatrix} 
%     \begin{bmatrix}
%         D_1 \\[4pt] D_2
%     \end{bmatrix}=   \begin{bmatrix}
%         A_3(\tau_p) & B_3(\tau_p)\\[4pt]
%         A_{3,\tau}(\tau_p) & B_{3,\tau}(\tau_p) 
%     \end{bmatrix} 
%     \begin{bmatrix}
%         D_3 \\[4pt]D_4
%     \end{bmatrix},
% \end{align}
where
\begin{align}
    A_{j}(\tau)= \frac{Ai(\zeta_j(\tau))}{\sqrt{\zeta_{j,\tau}(\tau)}}  \quad \mbox{and} \quad B_{j}(\tau)= \frac{Bi(\zeta_j(\tau))}{\sqrt{\zeta_{j,\tau}(\tau)}} \quad \mbox{for} \quad j=2,\,3.
\end{align}
Finally, the MAF solution of \eqref{eqn:eta_simplified_ver} over one period (and thus capturing both first and second turning points) is given by
% \begin{align}\label{eq:MAF_final}
%     \eta(\tau) =&\pi  \frac{\sqrt{\zeta_{,\tau}(0)}}{\sqrt{\zeta_{,\tau}(\tau)}}\begin{bmatrix}
%         Bi_{,\tau}\left(\zeta(0)\right) \\[6pt]
%        - Ai_{,\tau}\left(\zeta(0)\right)
%     \end{bmatrix}^T \begin{bmatrix}
%         Ai(\zeta(\tau)) \\[6pt]
%         Bi(\zeta(\tau))
%     \end{bmatrix} \\
%      &- \frac{\pi}{\sqrt{\zeta_{,\tau}(0)\zeta_{,\tau}(\tau)}} \left( \eta_{,\tau}(0)+ \eta(0) \frac{\zeta_{,\tau \tau}(0)}{2\zeta_{,\tau}(0)} \right) \begin{bmatrix}
%         Bi\left(\zeta(0)\right) \\[6pt]
%        - Ai\left(\zeta(0)\right)
%     \end{bmatrix}^T \begin{bmatrix}
%         Ai(\zeta(\tau)) \\[6pt]
%         Bi(\zeta(\tau))
%     \end{bmatrix}. 
% \end{align}
% \begin{align}
%     \eta(\tau)=\left\{ 
%     \begin{aligned}
%    & D_1 Ai\left(\zeta_1(\tau)\right) + D_2(\tau) Bi\left((\zeta_1(\tau))\right)/\sqrt{\zeta_{1,\tau}(\tau)} \quad \mbox{for} \quad 0\leq \tau \leq \tau_p \\
%     &D_3(\tau) Ai\left(\zeta_2(\tau)\right) + D_4(\tau) Bi\left((\zeta_2(\tau))\right)/\sqrt{\zeta_{2,\tau}(\tau)} \quad \mbox{for} \quad \tau_p \leq \tau \leq 2\pi.
%     \end{aligned}
%     \right. 
% \end{align}
\begin{align}\label{eq:MAF_final}
    \eta(\tau)=\left\{ 
    \begin{aligned}
   & \left(\zeta_{1,\tau}(\tau)\right)^{-1/2} \big[D_1 Ai\left(\zeta_1(\tau)\right) + D_2 Bi\left(\zeta_1(\tau)\right) \big] \quad \mbox{for} \quad 0\leq \tau \leq \tau_p, \\
    &\left(\zeta_{2,\tau}(\tau)\right)^{-1/2}\big[D_3 Ai\left(\zeta_2(\tau)\right) + D_4 Bi\left(\zeta_2(\tau)\right)\big]\quad \mbox{for} \quad \tau_p \leq \tau \leq 2\pi.
    \end{aligned}
    \right. 
\end{align}

\begin{figure}
    \centering
    \includegraphics[width=0.7\linewidth]{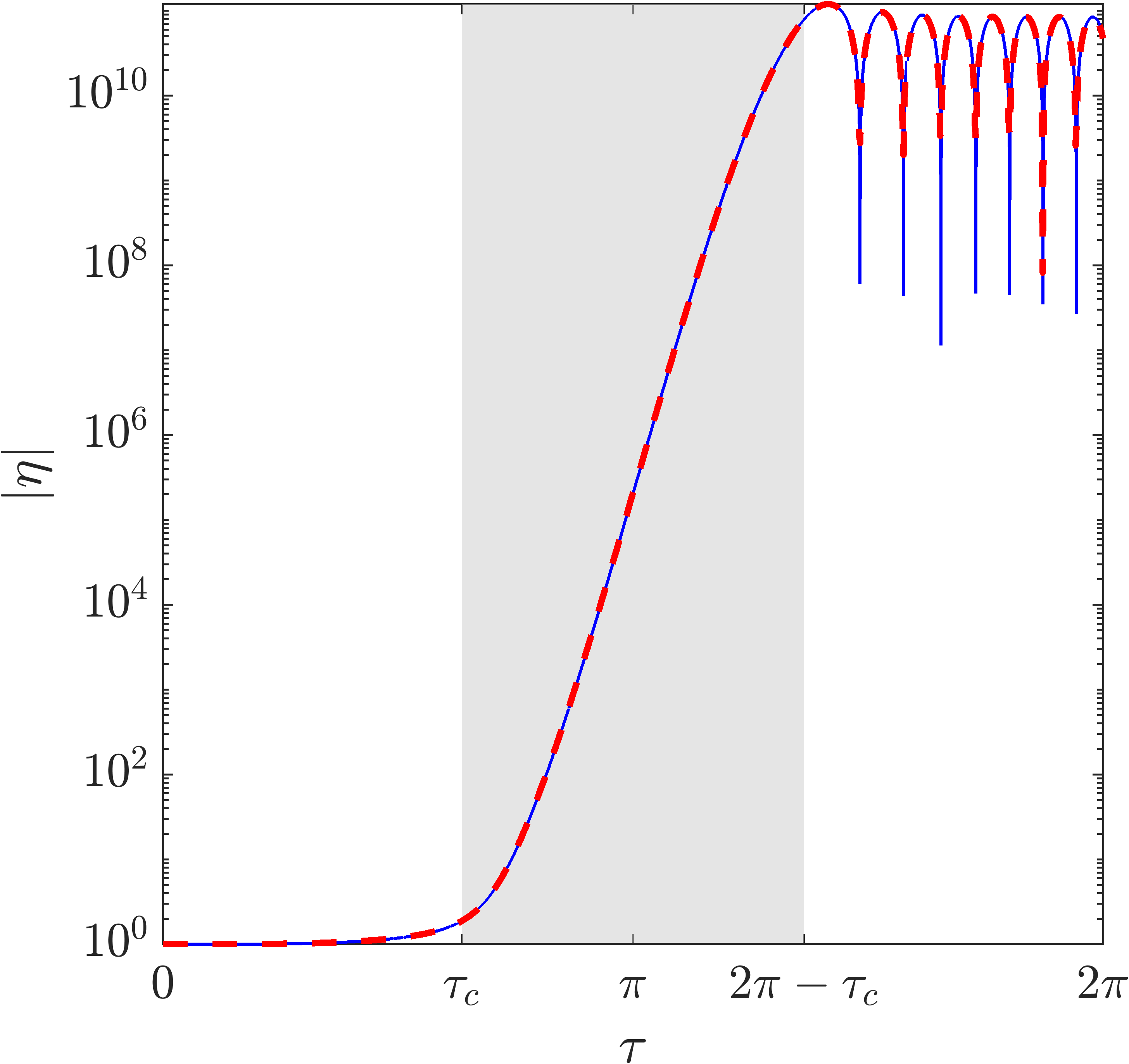}
    \caption{Comparison between numerical solution (blue) and MAF (dashed-red) for one period for the case $\beta=0.5$. The first and second turning points are respectively $\tau_c$ and $2\pi-\tau_c$;  the unstable region predicted from \eqref{eq:tau_c} is shaded in grey.}
    \label{fig:maf_full_period}
\end{figure}

{Figure\,\ref{fig:maf_full_period}, which is for the case $\beta\!=\!0.5$, reveals that the MAF solution is nearly indistinguishable from the numerical solution. Unlike figure\,\ref{fig:beta_p5}, figure\,\ref{fig:maf_full_period}  has been plotted for the entire $2\pi$ period and thus includes the behaviour around both turning points. While figure\,\ref{fig:maf_full_period} reveals oscillations after the second turning point, the existence of such a behaviour is contentious in reality since, even before the second turning point is reached, the flow is likely to enter the fully nonlinear stages of KH instability, making the linear solution invalid.}

\end{subsection}

\section{Numerical simulation: unsteady vortex blob method}
\label{sec:numer}
Here we extend the linear interfacial dynamics to the fully nonlinear stages via the vortex blob method, which is a standard numerical technique for capturing the nonlinear evolution of vortex sheets \citep{baker1982generalized,sohn2010long,pozrikidis2011introduction,bhardwaj2018nonlinear}. The applicability of the vortex blob method to our two-layered oscillating flow problem is fully justified since the two-fluid interface is a vortex sheet, i.e.\,the tangential component of fluid velocity has a jump discontinuity. In vortex methods, the interface is described as a parametric curve $\mathbf{x}(s,t)=(x(s,t),z(s,t))$ with $s$ being the arc-length parameter, and evolves according to
\begin{equation}
    \mathbf{x}_{,t}=\mathbf{q},
\end{equation}
where $\mathbf{q}\,=\,({\overline{u}},\,{\overline{w}})\,=\,(\mathbf{u_1}+\mathbf{u_2})/2$ is the arithmetic mean of the velocities above (i.e.\,$\mathbf{u_1}\!=\!(u_1,\,w_1$)) and below (i.e.\,$\mathbf{u_2}\!=\!(u_2,\,w_2$)) the interface. For a domain that is periodic in the along-channel direction $x$, the velocity $\mathbf{q}$ is obtained via the Birkhoff-Rott equation
\begin{equation}
\overline{u}-\mathrm{i} \overline{w} = \frac{\mathrm{i}}{2\lambda}\int_{0}^{\lambda}\widetilde{\gamma}\cot\bigg[\frac{\pi(\chi - \widetilde{\chi})}{\lambda}\bigg] d\widetilde{s},
\label{eq:BR_eq}
\end{equation}
where  $ \chi $ is the complex position of the interface: $ \chi = x(s,t) + \mathrm{i} z(s,t) $, $ \lambda $ is the wavelength, and $\widetilde{\gamma}\!=\!\gamma (\widetilde{s},t)$ denotes the vortex sheet strength, which is defined as the  jump in the tangential velocities of the two fluids across the
interface:
\begin{equation}
    \gamma=(\mathbf{u_1}-\mathbf{u_2})\cdot \hat{\mathbf{s}}.
    \label{eq:vortex_sheet_strn}
\end{equation}

In order to solve \eqref{eq:BR_eq}, the evolution equation for the vortex sheet strength is needed. To obtain this, first  we rewrite \eqref{eqn:main_equation_u}--\eqref{eqn:main_equation_w} for each layer in a vectorial form
\begin{equation}
    \frac{\mathrm{D}\mathbf{u_i}}{\mathrm{D}t} = -\frac{1}{\rho_0}\nabla P_{i} - g\frac{\rho_{i}}{\rho_0}\big(\sin \alpha (t) \hat{\mathbf{x}} + 
    \cos \alpha (t) \hat{\mathbf{z}} \big)+ 2 \alpha_{,t} \hat{\mathbf{y}} \times \mathbf{u_i},
    \label{eq:vectorial}
\end{equation}
where $\mathbf{u_i}\!=\!(u_i,w_i)$, and $i\!=\!1(2)$ implies upper (lower) layer.  Subtracting the equation for $i\!=\!2$ from $i\!=\!1$ and projecting the resultant equation along the tangential direction (by taking an inner product  with $\mathbf{\hat{s}}$)  yields
\begin{equation}
    \frac{D\gamma}{D t}+\gamma\frac{\partial \mathbf{q}}{\partial s}\cdot \mathbf{\hat{s}}=g'\big(\sin \alpha (t) \hat{\mathbf{x}} + 
    \cos \alpha (t) \hat{\mathbf{z}} \big)\cdot \mathbf{\hat{s}}+\bigg[2 \alpha_{,t} \hat{\mathbf{y}} \times (\mathbf{u_1}-\mathbf{u_2})\bigg]\cdot \mathbf{\hat{s}},
    \label{eq:vortex_sheet_evol_eq}
\end{equation}
where $D/Dt\!=\!\partial/\partial t + \mathbf{q} \cdot \mathbf{\hat{s}}\, \partial/\partial s$. In obtaining \eqref{eq:vortex_sheet_evol_eq}, the continuity of pressure across the interface has been used. Furthermore, using the identities $\hat{\mathbf{x}}\cdot \hat{\mathbf{s}}=\partial x/\partial s$ and $\hat{\mathbf{z}}\cdot \hat{\mathbf{s}}=\partial z/\partial s$, we find that the last term in the RHS of \eqref{eq:vortex_sheet_evol_eq}, arising from the Coriolis acceleration, vanishes since the interface is a streamline:
\begin{equation*}
    (\hat{\mathbf{y}} \times \mathbf{u_{1,2}} )\cdot \hat{\mathbf{s}}=w_{1,2} \frac{\partial x}{\partial s}-u_{1,2}\frac{\partial z}{\partial s}=0.
\end{equation*}
%$(\hat{\mathbf{y}} \times \mathbf{u_{1,2}} )\cdot \hat{\mathbf{s}}=w_{1,2} \partial x/\partial s-u_{1,2}\partial z/\partial s=0$.

Previous studies with vortex methods have primarily been restricted to steady background flows (for example, see \cite{sohn2010long} and references therein)  although the method itself applies to unsteady flows. In our case, the explicit time dependence of the flow clearly manifests in the RHS of \eqref{eq:vortex_sheet_evol_eq} through the angle of oscillation parameter $\alpha (t)$. 

For the numerical implementation, we first write \eqref{eq:BR_eq} in a discretised fashion:
\begin{subequations}
\begin{align}
 \overline{u}_m &= x_{m,t} = \frac{1}{2}\sum_{n=1, n \neq m}^{N} {\Gamma_{n}}\frac{\sinh [k(z_{m}-z_{n})]}{\cosh [k(z_{m}-z_{n})] - \cos [k(z_{m}-z_{n})] + \delta^2}, \label{eq:birkhoff_disc_u}\\
\overline{w}_m&= z_{m,t} = -\frac{1}{2}\sum_{n=1, n \neq m}^{N}{\Gamma_{n}}\frac{\sin[ k(x_{m}-x_{n})]}{\cosh[ k(z_{m}-z_{n})] - \cos[ k(z_{m}-z_{n})] + \delta^2}, \label{eq:birkhoff_disc_w}  
\end{align}
\end{subequations}
where $(\overline{u}_m,\overline{w}_m)$ is the velocity of the $m^{\mathrm{th}}$  point vortex (Lagrangian marker) at the location ($x_m,z_m$), $N$ denotes the total number of point vortices,  $k\!=\!2\pi/\lambda$ denotes the wavenumber, and $\delta$ denotes  Kransy's parameter \citep{krasny1986desingularization} which regularises the otherwise singular kernel in \eqref{eq:BR_eq}. The parameter $\delta$ can be regarded as a proxy for viscous effects, numerically replacing point vortices with vortex `blobs', and thereby adding a finite thickness to the otherwise discontinuous shear layer. 
Finally, $\Gamma_m$ denotes the circulation of the $m^{\mathrm{th}}$ point vortex, which is given as follows: 
\begin{equation}
\Gamma_{m} = \gamma_{m} \Delta s_{m},
\label{eq:Gamma_eq}
\end{equation}
where $ \Delta s_{m} $ is the arc-length about the $m^{\mathrm{th}}$ point  vortex:
\begin{equation*}
\Delta s_{m} = \sqrt{ \Delta x_m^2 + \Delta z_m^2}.
\end{equation*}
In the above equation, $\Delta x_m=(x_{m+1} - x_{m-1})/2$ and $\Delta z_m=(z_{m+1} - z_{m-1})/2$. Equations \eqref{eq:birkhoff_disc_u}--\eqref{eq:birkhoff_disc_w} succinctly show that the interaction between the point vortices representing the discretised vortex sheet leads to interface evolution. 

Using \eqref{eq:Gamma_eq} in \eqref{eq:vortex_sheet_evol_eq}, we obtain the discretised version of the circulation evolution equation:
\begin{equation}
    \Gamma_{m, t}  = \Delta x_m \,g'\sin{\alpha (t)}  + \Delta z_m\, g'\cos{\alpha (t)} .
    \label{eq:evol_circ}
\end{equation}

Next, we numerically solve \eqref{eq:birkhoff_disc_u}--\eqref{eq:birkhoff_disc_w} along with \eqref{eq:evol_circ} (where the definition of $\alpha (t)$ is given in \eqref{eqn:alpha}) using the fourth-order Runge-Kutta time integration scheme. To speed up the computation, we recast the governing equations into the slow-time variable $\tau$.  The system is  initialised with a short, linear, interfacial gravity wave with the following interfacial displacement, written in a discretised fashion:   $z_m\!=\!a\cos(kx_m)$, where $x_m\!=\!m/N$, $k\!=\!2\pi$ and $a\!=\!10^{-6}$. The initial circulation of the $m^{\mathrm{th}}$ point vortex is then given by $$\Gamma_m(0)=-2\omega z_m \Delta x_m,$$ 
where $\omega\!=\sqrt{g'k/2}\!=\!0.1772$, and corresponds to $g'\!=\!0.01$. Initially we take $N\!=\!400$.  With the progress of time, especially during the highly nonlinear stages, the interface can develop poor resolution in certain locations. To tackle this, we follow the vortex insertion scheme outlined in \citet{sohn2010long}. In this procedure,  a threshold arc-length $ \Delta s_{\mathrm{thresh}} $ is defined; if the arc-length at any location $ m $ becomes greater than the threshold, i.e.\ $ \Delta s_{m}\!>\!\Delta s_{\mathrm{thresh}} $, then a point vortex is inserted. The $x$ coordinate of the inserted vortex is taken to be the average of the $x$ coordinates of $m^{\mathrm{th}}$ and $(m+1)^{\mathrm{th}}$ vortices; the same approach is followed for the $z$ coordinate and the circulation. The Kransy's parameter $\delta$ is chosen accordingly to suppress spurious high wavenumber instabilities. 
%Following \citet{sohn2010long}, the Kransy's parameter is fixed at $\delta=0.1$ in all simulations. 

\begin{figure}
    \centering
    \begin{subfigure}{0.48\textwidth}
        \includegraphics[width=0.95\linewidth]{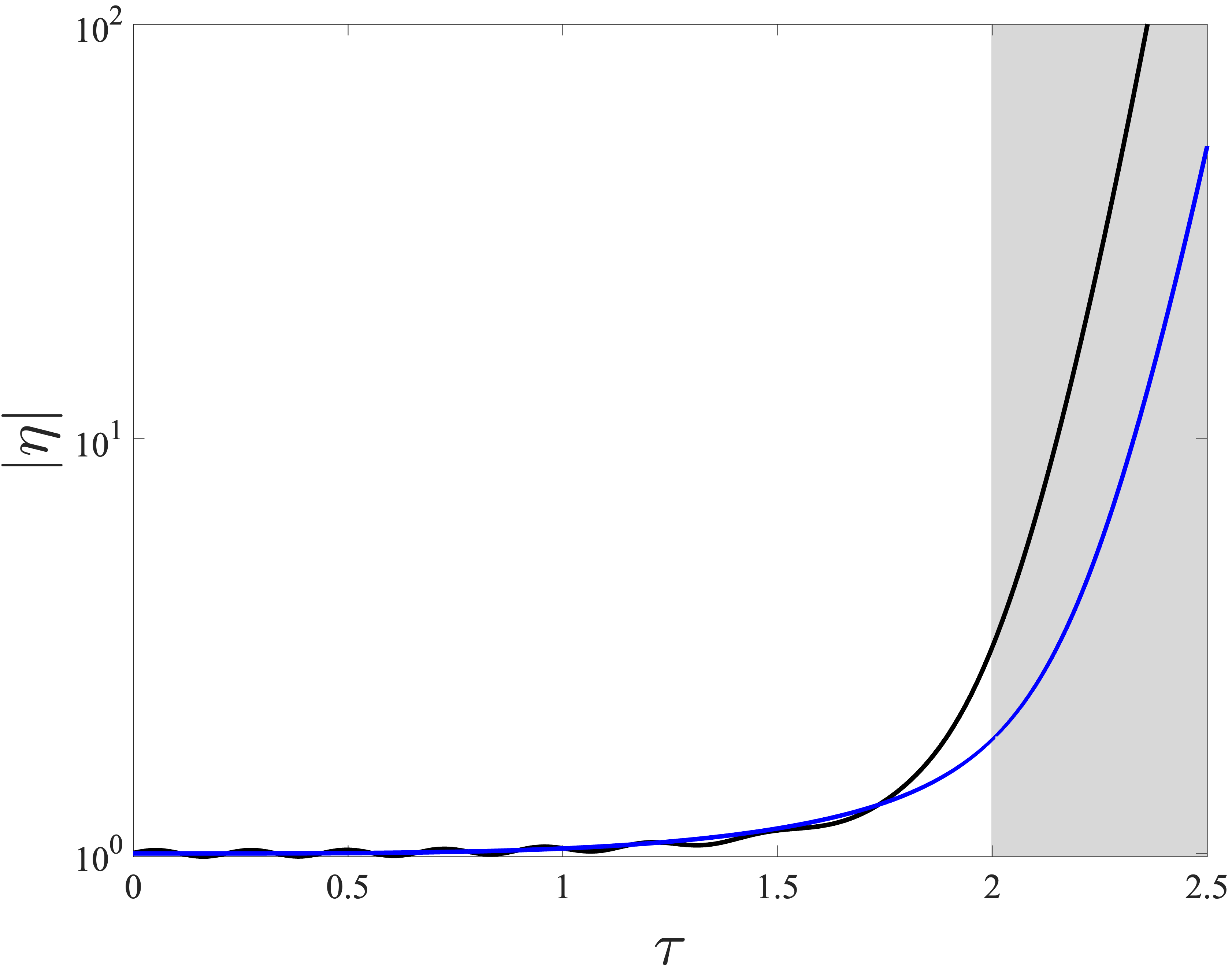}
    \caption{$\beta=0.5$}
    \label{fig:beta_num_050}
\end{subfigure}
%\hfill
\begin{subfigure}{0.48\textwidth}
        \includegraphics[width=0.95\linewidth]{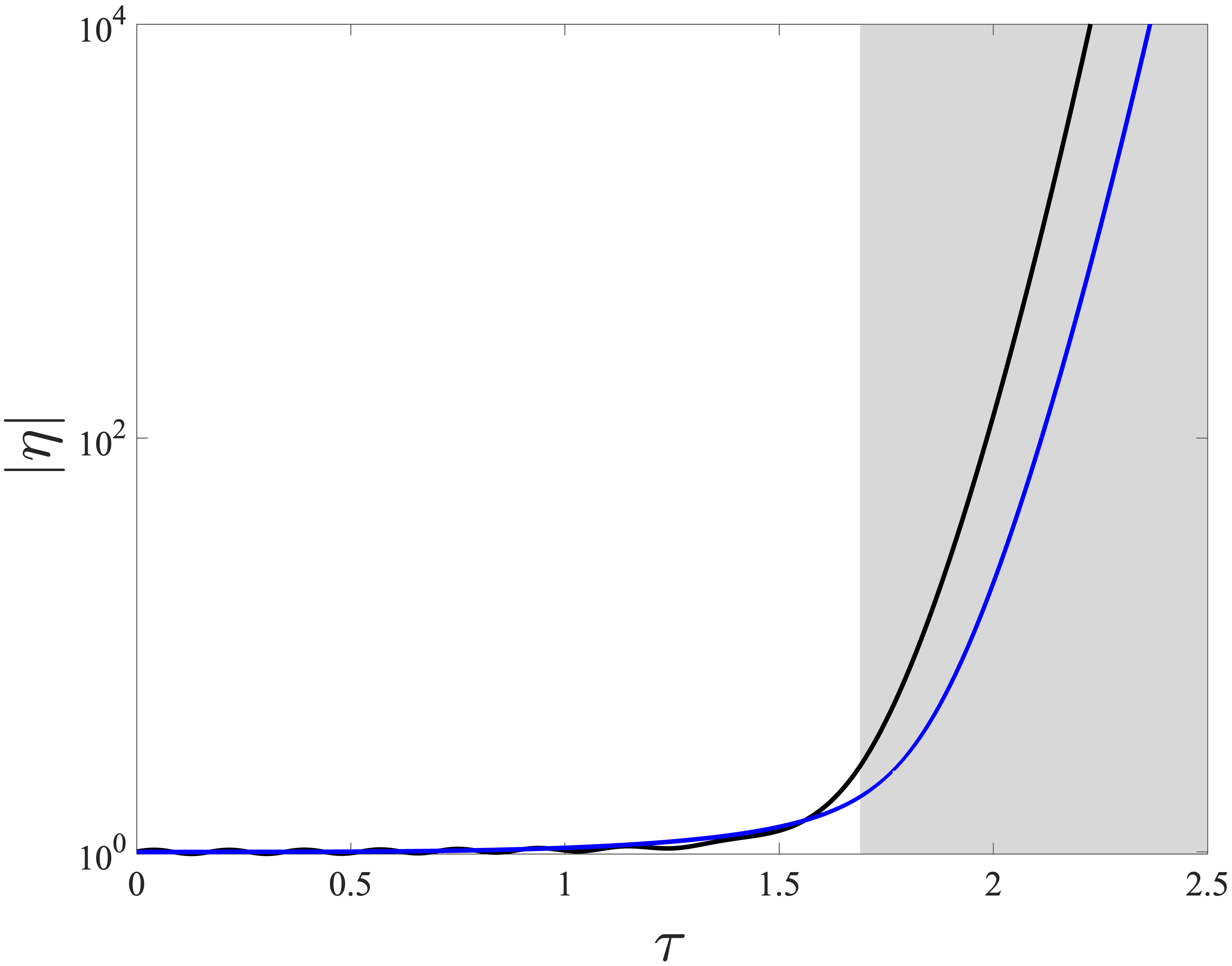}
    \caption{$\beta=0.8$}
    \label{fig:beta_num_080}
\end{subfigure}
\caption{Interface amplitude evolution for different $\beta$ values. Blue - solution to the linear equation \eqref{eqn:eta_simplified_ver}, black -  solution obtained using the vortex method. Unstable region predicted from \eqref{eq:tau_c} is shaded in grey.}  
\label{fig:comaprisons_with_VM}
\end{figure}

\begin{figure}
    \centering
    \includegraphics[scale=0.3]{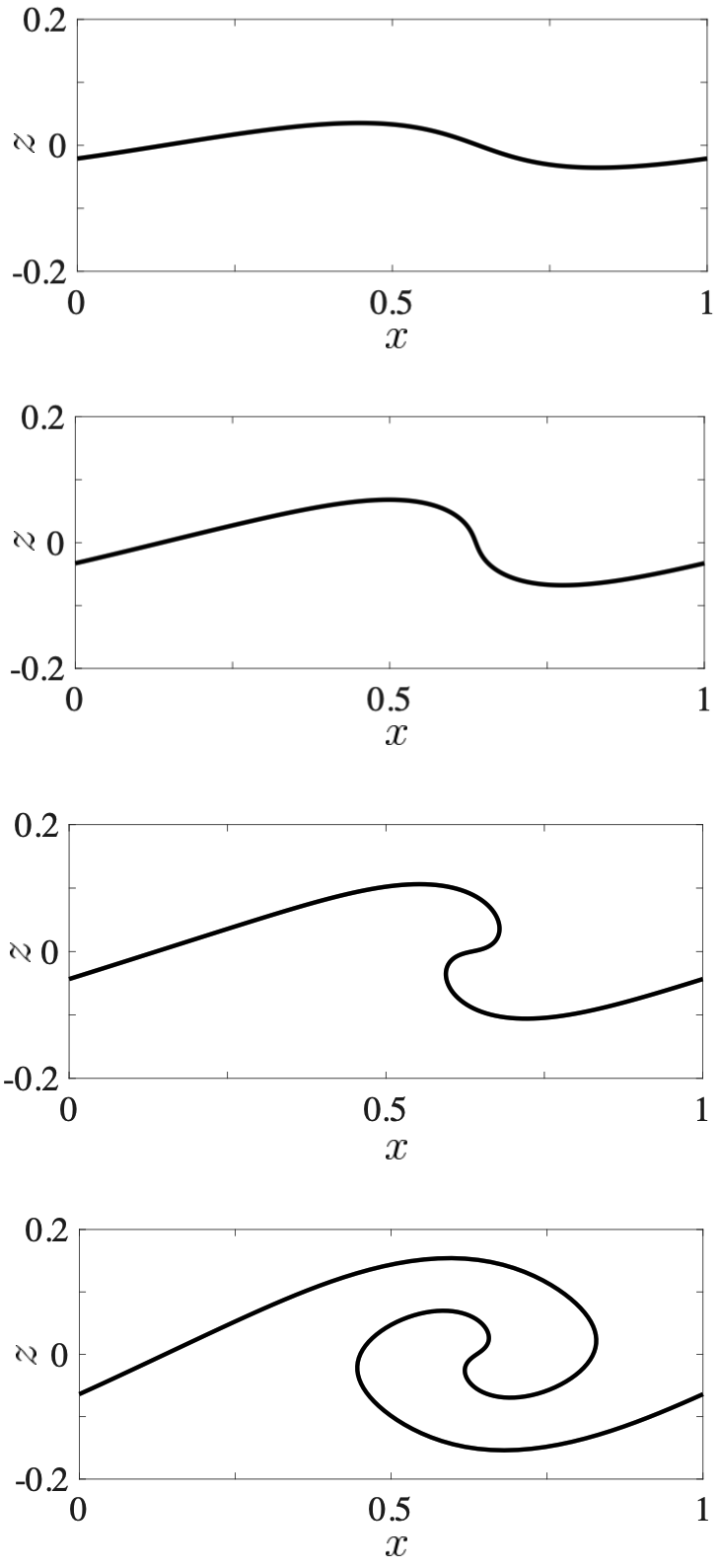}
    \caption{Interface evolution for the case $\beta=0.8$ for $\tau =2.285, 2.315, 2.340$ and $2.395$.}
    \label{fig:beta_08_evol}
\end{figure}

Figure \ref{fig:comaprisons_with_VM} compares the interface evolution obtained from the linear equation  \eqref{eqn:eta_simplified_ver} with the fully nonlinear solution obtained using the vortex blob method for two different $\beta$ values. The growth rate and the onset of instability predicted from the vortex method solution agree well with the linear theory.

Figure \ref{fig:beta_08_evol} shows the interface evolution for the case $\beta=0.8$. The interface rolls up into a KH billow, qualitatively agreeing with the DNS results of \cite{inoue2009efficiency} and \cite{lewin2022stratified}. The numerical simulation is run until the billow saturates. 
 Any further late-time dynamics may not be representative of the reality since the vortex method is strictly 2D and hence does not capture the 3D effects emanating from secondary instabilities, which can crucially alter the dynamical behaviour \citep{lewin2022stratified}.

\section{Physical interpretation and case studies}
\label{bsec:case_study}
The oscillating tank system is a simplified model for studying instabilities arising in a temporally varying pycnocline in lakes, estuaries, and oceans.  Pycnocline oscillations are typically generated by wind forcing, or by diurnal and semi-diurnal tides interacting with underwater topography \citep{sanford1990,mihanovic2009diurnal}. 
During the summer months, many shelf seas and lakes reveal a two-layered density stratification \citep{lemmin2005internal,mihanovic2009diurnal,guihou2018kilometric}, providing ideal conditions to compare the flow field induced by low baroclinic mode oscillations of the pycnocline with the oscillating tank system. Below we provide two case studies, one for Lake Geneva and the other for the Chesapeake Bay. 

\subsection{Lake Geneva}
\label{subsec:geneva}
We consider the background conditions of the summer of $1950$ given in \cite{lemmin2005internal}: $h_1\!=\!15$ m, $h_2\!=\!175$ m, $\rho_1\!=\!998.49$ kg/m$^3$ (corresponding to $19^\circ$C) and  $\rho_2\!=\!1000$ kg/m$^3$ (corresponding to $5.5^\circ$C). The frequency of mode-1 is $\omega_f\!=\!2.14 \times 10^{-5}$ s$^{-1}$ (period of $81.5$ hours) and corresponds to a Kelvin seiche. The maximum width of the lake is $13.8$ km and the maximum difference in thermocline excursions at the two extremities of the lake-width (station-3 and station-8) due to mode-1 is $\approx\!2$ m. Hence $\tan \alpha_f \approx \alpha_f = 2/13,800=0.000145$. Since the two layers are of unequal depths, the analysis provided in Appendix \ref{appA} will be applicable. The problem can be simplified since the lower layer is much deeper than the upper layer, which leads to the instability condition $\beta\!>\!1/8$;   see \eqref{eq:beta_unequal_depth}. This yields
\begin{align}
  \beta=\alpha_f^2\frac{ \omega^2}{\omega_f^2}>\frac{1}{8} \implies k > \frac{\omega_f^2}{4g'\alpha_f^2}\,\,\,\,\, \mathrm{or}\,\,\,\,\, \lambda< \frac{8\pi g'\alpha_f^2}{\omega_f^2}.
    \label{eq:lambda_max_lake}
\end{align}
For the parameters of Lake Geneva, the above equation would imply that interfacial waves at the pycnocline shorter than $\lambda_{\mathrm{long-wave-cut-off}}=17.03$ m will be rendered unstable.   An interfacial wave of wavelength $17$ m will start to grow after $\tau_c=\cos^{-1}[1-(2\beta)^{-1/2}]= 40$ hours (see  Appendix \ref{appA}) from initiation of the mode-1 oscillation. 
%However, it is important to note that these values should be taken as a rough estimate since the value of $0.66$ m for the longest unstable wavelength  is too small to discard the finite thickness effects of the pycnocline (pycnoclines always have a finite thickness, however small). Viscous effects can also start to play a role at these scales, which have been ignored in our inviscid analysis.}
%(since these predictions stem from the $$ from our discontinuous setting is verified against the smooth profile result of \cite{lewin2022stratified}). Furthermore, the long-wave cut-off, while might not be accurate, can provide an estimate of the actual wavelength in question.}

\subsection{Chesapeake Bay}
\label{subsec:chesapeake}
Here we consider the background conditions of May/June $1993$ of the Chesapeake Bay given in \cite{frizzell1997temporal}. In this case, $h_1\!=\!10$ m, $h_2\!=\!13$ m, $\rho_1\!=\!1004$ kg/m$^3$, and $\rho_2\!=\!1011$ kg/m$^3$. Pycnocline oscillations had a period of $12$ hours ($\omega_f\!=\!0.00014$ s$^{-1}$), which was generated by the interaction of semi-diurnal tides with a discontinuity in the bathymetry. About $4$ m of vertical displacement over half of the wavelength ($0.5\times 25$ km $=12.5$ km) of the internal tide was observed, yielding $\tan \alpha_f \approx \alpha_f = 4/12,500=0.00032$. The two layers have unequal depths with $r\!=\!h_1/h\!=\!0.435$, yielding the instability condition $\beta\!>\!0.246$. Following the analysis in Appendix \ref{appA}, we obtain
\begin{align}
  \beta=\alpha_f^2\frac{ \omega^2}{\omega_f^2}>0.246 \implies k > \frac{0.492 \omega_f^2}{g'\alpha_f^2}\,\,\,\,\, \mathrm{or}\,\,\,\,\, \lambda< \frac{4.065\pi g'\alpha_f^2}{\omega_f^2}.
    \label{eq:lambda_max_bay}
\end{align}
Hence, for the parameters of Chesapeake Bay, interfacial waves at the pycnocline shorter than 
$\lambda_{\mathrm{long-wave-cut-off}}=4.2$ m will become unstable. This implies that an interfacial wave of wavelength $4$ m would start to grow  $\tau_c=5.4$ hours (see \eqref{eq:tau_c_general}) after the inception of the internal tide. However, it is important to note that these values should be taken as a rough estimate since the value of $4.2$ m for the longest unstable wavelength is not large enough to disregard the finite thickness effects of the pycnocline (pycnocline always has a finite thickness, however small). 

\section{Summary and conclusion}
\label{sec:disc_conc}
This study uses mathematical and numerical techniques to investigate the generation mechanism and fundamental characteristics of instabilities arising at the density interface in a two-layered oscillating channel flow. The setup provides a simplified model for studying shear instabilities arising at the density interface due to the time-periodic shear generated by the slow oscillations of the density interface itself. This self-induced shear of the oscillating pycnocline may provide an alternate pathway to turbulence and diapycnal mixing in addition to that obtained via wind-induced shear in the upper ocean or the lake epilimnion.
%instabilities arising in short interfacial gravity waves due to the shear produced by long interfacial waves in various geophysical settings, such as lakes and oceans. This wave-driven shear instability leads to an alternate route to small-scale turbulence and diapycnal mixing, and might be especially important in deep oceans \citep{van2010deep}, or in the upper ocean (as well as the lake epilimnion) when the wind-induced shear is weak. Seiches in stratified, long, narrow lakes could also generate time-period shear, and could potentially lead to such short-wave instabilities.

The equation governing the density interface dynamics, given by \eqref{eqn:eta_simplified_ver}, is a Schr\"{o}dinger-type equation with a time-periodic potential.
%Explicit time dependence (periodic to be specific) is the hallmark of the problem.
The small parameter $\epsilon$, which is the ratio of the low-frequency pycnocline oscillations (forcing frequency) and the high-frequency short waves (that could be rendered unstable), encapsulates the separation of time scales. This makes \eqref{eqn:eta_simplified_ver} fundamentally different from the commonly studied time-periodic dynamical systems governed by Mathieu-type equations, which can lead to parametric instabilities. An inverse Richardson number like nondimensional parameter $\beta$  determines the necessary and sufficient stability condition of \eqref{eqn:eta_simplified_ver}; for two layers of equal thickness, the flow becomes linearly unstable when $\beta\!>\!1/4$. When the layers have unequal thickness, the stability condition depends on the ratio of layer thicknesses, see Appendix \ref{appA}.  Once the instability condition is satisfied, the solution does not start to grow after $\tau\!=\!0+$, but rather tunnels into the unstable state after $\tau\!=\!\tau_c$. Using the parameters of a DNS case study in \cite{lewin2022stratified}, we show that our theory correctly predicts the time of onset of instability. Next, we study the interface evolution by numerically solving \eqref{eqn:eta_simplified_ver} and compare it with the asymptotic solution obtained via the MAF method.  We show that MAF provides a uniformly valid composite solution over the entire period (which includes two turning points) and reveals an excellent agreement with the numerical solution. To evaluate the temporal growth rate, we compute the Floquet exponents; furthermore, we also derive the exact normal mode growth rate obtained for various steady-state-based approximations of the background flow. We observe that the normal mode theory implemented on a background flow that considers the maximum velocity in each layer (as opposed to the time average over a period) provides a highly accurate growth rate prediction. The normal mode growth rate is shown to exactly match the classical vortex-sheet KH instability, thereby conclusively proving that the instability is of KH type.
Next, we extend the linear analysis of the interface evolution to the fully nonlinear stages via the unsteady vortex blob method. While the standard vortex blob method is commonly implemented for steady background flows, the procedure also applies to time-dependent flows. This needs modification to the circulation evolution equation typically used in vortex methods problems with density stratification. We derive the unsteady circulation evolution equation \eqref{eq:evol_circ}, which captures the explicit time dependence of the flow. The time of the onset of instability as well as the growth rate show good agreement with the predictions from the linear analysis. During the nonlinear stages, the interface rolls up into KH billows, agreeing with the previous DNS studies \citep{inoue2009efficiency,lewin2022stratified}.

Finally, we provide a physical context to the two-layered oscillating tank model by considering two case studies -- Lake Geneva and Chesapeake Bay. Each of these sites reveals pycnocline oscillations, and using the available data, we obtain the stability condition, estimate the time of onset of the shear instabilities as well as the long-wavelength cut-off.

As a closing remark, we reiterate the fact that stability analysis with discontinuous profiles provides spurious growth rates (ultraviolet catastrophe), which necessitates stability analysis of background flows with finite shear layer thickness. The latter is also essential for obtaining the wavenumber corresponding to the maximum growth rate, {as well as a short-wavelength cut-off}, which cannot be obtained from discontinuous analysis. Despite these limitations, the two-layered configuration studied in this paper provides a foundational basis for the unsteady shear instability mechanism, makes the problem analytically (or semi-analytically) tractable through a simple governing equation, and also provides the necessary and sufficient condition for instability, which leads to key predictions (e.g.\, the time of onset of instability) that excellently compare with numerical results obtained from smooth profiles.

\section*{Acknowledgment}
We sincerely thank Dr.\,Sam Lewin from the University of California, Berkeley and the two anonymous reviewers for the helpful comments and suggestions. We are grateful to Dr.\,Sam Lewin for running the DNS, which allowed us to compare the $\tau_c$ value from our predictions with the DNS results. We also thank the Associate Editor, Prof. Neil Balmforth, from the University of British Columbia, for his insightful comments.  AG thanks Dr.\,Jeff Polton from the National Oceanographic Centre, Liverpool for fruitful discussions  
on oceanographic applications of the problem.

\section*{Declaration of interests}
The authors report no conflict of interest. 

\appendix
\section{Derivation of the stability condition for two layers of unequal depths}\label{appA}
Let us consider the two layers having different depths, denoted by $h_1$ and $h_2$.  The background state density distribution is then given by
\begin{align}\label{eqn:rho_unequal_depth}
    \rho(z) =
    \left\{
    \begin{aligned}
    & \rho_1,~~~ 0<z<h_1, \\
    & \rho_2,~~~ -h_2<z<0,
    \end{aligned}
    \right.
\end{align}
where  $h_2\!=\!h-h_1$, in which $h$ represents the total channel depth. 
%represents the total depth of the rectangular cross section of the tank and $\rho_1 < \rho_2$.
Following the method outlined in \S \ref{subsec:bg_flow}, the  background state velocity profile is given by:
\begin{align}\label{eq:vel_unequal_depth}
    U(z,t)=\left\{
    \begin{aligned}
    & ~~\frac{g' h_2}{h} \int_0^t\sin\alpha(\tilde{t})\, \mathrm{d}\tilde{t}, ~~~ 0<z<h_1 \\
    &-\frac{g' h_1}{h} \int_0^t\sin\alpha(\tilde{t})\, \mathrm{d}\tilde{t},~~~ 
    -h_2<z<0.
    \end{aligned}
    \right.
\end{align}
Under the assumption of small oscillations, the background state velocity profile simplifies to
\begin{align}\label{eq:vel_unequal_depth}
    U(z,t)=\left\{
    \begin{aligned}
    & ~~\frac{g' h_2}{h}\frac{\alpha_f}{\omega_f} \bigg(1-\cos\left(\omega_f t\right)\bigg),~~~ 0<z<h_1 \\
    &-\frac{g' h_1}{h} \frac{\alpha_f}{\omega_f} \bigg(1-\cos\left(\omega_f t\right)\bigg),~~~ 
    -h_2<z<0.
    \end{aligned}
    \right.
\end{align}

Following the procedure in \S \ref{subsec:lin_per}, infinitesimal perturbations are added to the background state. The waves are assumed to be significantly shorter than the depth of the individual layers, i.e. $kh_1\!\gg\!1$ and $kh_2\!\gg\!1$. The evolution equation of the interface $\eta (\tau)$ between the two fluid layers under the Boussinesq approximation is given by
% \begin{align}
%     \epsilon^2\eta_{,\tau \tau} +  \left[ 1- \frac{\alpha_f^2}{4} (1-\cos(2\omega_f t))  - 2\beta^2  \frac{(h_1^2+h_2^2)}{ h^2}\left(\frac{3}{2} -2\cos(\omega_f t) +\frac{1}{2}\cos(2\omega_f t)\right) \right]\eta = 0
% \end{align}
% \begin{align}
%      \epsilon^2\eta_{,\tau \tau} + \left[ 1 - 3\beta \frac{h_1^2+h_2^2}{h^2} + 2\beta \frac{h_1^2+h_2^2}{h^2} \left( 2\cos(\tau) -\frac{1}{2}\cos(2\tau)\right) + \mathcal{O}(\alpha_f^2)\right] \eta =0.
% \end{align}
\begin{align}
     \epsilon^2\eta_{,\tau \tau} + \bigg[ 1 - 3\beta \left(r^2+(1-r)^2\right)  +  2\beta \left(r^2+(1-r)^2\right) \left( 2\cos(\tau) -\frac{1}{2}\cos(2\tau)\right)  \nonumber \\ + \mathcal{O}(\alpha_f^2)\bigg] \eta =0,
\end{align}
where  $r\!=\!h_1/h$. All variables and parameters are the same as in \S \ref{subsec:lin_per}. 
%Note that in the equation above, we have applied the Boussinesq approximation and assumed that the waves are significantly shorter than the thickness of the individual layers.  
The necessary and sufficient condition for instability is found to be
\begin{align}
    \beta > \beta_{\mathrm{min}}= \frac{1}{8\left[r^2+(1-r)^2\right]}.
    \label{eq:gen_inst_cond}
\end{align} 
Since $r\in(0,1)$, this implies that $1/8< \beta_{\mathrm{min}} \leq 1/4$.

The turning point $\tau_c$, marking the time at which the flow tunnels into the unstable region is given by
\begin{equation}
    \tau_c=\cos^{-1} \Bigg( 1-\frac{1}{\sqrt{2\beta\left[r^2+(1-r)^2\right]}}\Bigg).
    \label{eq:tau_c_general}
\end{equation}
The maximum growth rate obtained from the normal-mode analysis is as follows:
\begin{equation}
    \sigma_{\mathrm{normal-mode}}=\frac{\sqrt{8\beta\left[r^2+(1-r)^2\right]-1}}{\epsilon}.
\end{equation}
%$\sigma_{\mathrm{normal-mode}}=\sqrt{8\beta\left[r^2+(1-r)^2\right]-1}/\epsilon$.
In a typical oceanographic scenario, the depth of the pycnocline is much smaller than the total depth of the ocean, i.e. $r\!\ll\!1$.  In this situation, the instability condition becomes $\beta \!>\!1/8$, the instability will start after $\tau_c=\cos^{-1}[1-(2\beta)^{-1/2}]$, and the maximum growth rate obtained from the normal-mode method simplifies to $\sigma_{\mathrm{normal-mode}}=\sqrt{8\beta-1}/\epsilon$.

%\bibliographystyle{jfm}
% Note the spaces between the initials
\bibliography{jfm}

\end{document}

\maketitle

\vspace{-1cm}

\begin{center}
$^{1}$Department of Mathematics and Statistics, Gandhi Institute of Technology and Management (GITAM), Hyderabad 502329, India\\[4pt]
$^{2}$School of Science and Engineering, University of Dundee, Dundee DD1 4HN, UK
\end{center}

\vspace{1em}

% ---------- Reference to original article ----------
\noindent
% \textbf{Original article:}\\
% Author(s), \textit{Journal Name}, \textbf{Volume}, page--page (year).  
% DOI: \href{https://doi.org/xx.xxxx/xxxxx}{xx.xxxx/xxxxx}

%\vspace{1cm}

% ---------- Corrigendum text ----------
\noindent
This corrigendum corrects the generalised analysis (derived for an arbitrary $r\in(0,1)$) undertaken in Appendix~A of  \cite{Biswas_Guha_2025}. This error does not affect the vast majority of the analysis and conclusions of the paper, which are derived for the case $r=1/2$ (i.e. two fluid layers of equal depths).
%however, it changes the generalised equation (A4), which is derived for an arbitrary $r\in(0,1)$. 

%\bigskip

% ---------- Correction section ----------
%\noindent
%\textbf{Correction}

%\medskip

\noindent
In Appendix A (see page~21), the amplitude equation (A4) should read as
\begin{equation}
\begin{split}
   & \epsilon^2 \mathcal{N}_{,\tau \tau} +\bigg[1 -\frac{3}{2} \beta \left( r^2+(1-r)^2+\frac{1}{2}\right)+\\
   &\beta \left( r^2+(1-r)^2+\frac{1}{2}\right)\left(2 \cos(\tau) -\frac{1}{2}\cos(2\tau) \right) + \mathcal{O}(\alpha_f^2)\bigg]\mathcal{N} =0.
    \end{split}
\end{equation}
%instead of
% \begin{align}
%      \epsilon^2\eta_{,\tau \tau} + \bigg[ 1 - 3\beta \left(r^2+(1-r)^2\right)  +  2\beta \left(r^2+(1-r)^2\right) \left( 2\cos(\tau) -\frac{1}{2}\cos(2\tau)\right)   + \mathcal{O}(\alpha_f^2)\bigg] \eta =0,
% \end{align}
% where,
% \begin{equation}
% \eta(t)=\mathcal{N}(t)\,
% \exp\!\left[
% -\mathrm{i}\,\frac{k g'}{4}(1-2r)\frac{\alpha_f}{\omega_f}
% \left(
% t-\frac{\sin(\omega_f t)}{\omega_f}
% \right)
% \right].
% \end{equation}
\medskip
The necessary and sufficient condition for instability (see (A5)) should be corrected  to:
\begin{align}
    \beta >\beta_{min} = \frac{1}{4\left[r^2+(1-r)^2+\frac{1}{2} \right]}.
    \label{eq:2}
\end{align}
Since $r\in (0,1)$, we have $1/6 < \beta_{min} \leq 1/4$. The turning point $\tau_c$ in (A6) changes to 
\begin{align}
    \tau_c = \cos^{-1}\left(1-\frac{1}{\sqrt{\beta\left[r^2+(1-r)^2+\frac{1}{2} \right]}}\right).
    \label{eq:3}
\end{align}
% Normal mode growth rate is given by 
% \begin{align}
%     \sigma_{normal-mode} = \frac{1}{\epsilon}\left(\sqrt{4\beta\left[r^2+(1-r)^2+\frac{1}{2} \right]}-1\right)
% \end{align}

These lead to the following changes in the main body of the paper:

\begin{itemize}
\item \textbf{Section~2.2.1:} Equation (2.24) in the paper should be replaced by \eqref{eq:2}.
% re

% The necessary and sufficient condition for instability is corrected as
% \begin{align*}
%     \beta >\beta_{min} = \frac{1}{4\left[r^2+(1-r)^2+\frac{1}{2} \right]}
% \end{align*}
\item \textbf{Section~5.1:}
The instability condition is corrected to \(\beta > \tfrac{1}{6}\), and the
long-wave cut-off wavelength is corrected to
\(\lambda_{\text{long-wave cut-off}} = 12.77\,\mathrm{m}\) (instead of \(17.03\,\mathrm{m}\)).
The corresponding turning point time is given by
\[
\tau_c = \cos^{-1}\!\left(1 - (3\beta/2)^{-1/2}\right).
\]
The marginally unstable long wave starts to grow $\sim\!40$ hours (no change) from the inception of mode-1 oscillations with a time period of $\sim\!80$ hours.

\item \textbf{Section~5.2:}
The instability condition is corrected to \(\beta > 0.248\) (instead of \(\beta > 0.246\)), and the
long-wave cut-off wavelength is corrected to
\(\lambda_{\text{long-wave cut-off}} = 4.17\,\mathrm{m}\) (instead of \(4.2\,\mathrm{m}\)). As is obvious, the change here is insignificant.
\end{itemize}

%The error does not affect the conclusions of the paper. 

\bibliographystyle{apalike}
% Note the spaces between the initials
\bibliography{jfm}

% ---------- References (if needed) ----------
% \begin{thebibliography}{9}
% \bibitem{orig}
% Biswas L, Guha A. , \textit{Stability analysis of time-periodic shear flow generated by an oscillating density interface.}, J. Fluid Mech.,1015:A27. (2025).
% \end{thebibliography}